\documentstyle[11pt,aaspp4,tighten]{article}

\def\today{\rightline{\ifcase\month\or
        January\or February\or March\or April\or May\or June\or
        July\or August\or September\or October\or November\or December\fi
        \space\number\day, \number\year}}
\def\etal{{\it et al.}\ }

\def\msun{{\rm\,M_\odot}}

\def\s-1{{\rm\,s^{-1}}}

\def\spose#1{\hbox to 0pt{#1\hss}}
\def\C3H2{{\rm\,\rm C_3H_2}}
\def\NH3{{\rm\,\rm NH_3}}
\def\HOCO+{{\rm\,\rm HOCO^+}}
\def\lta{\mathrel{\spose{\lower 3pt\hbox{$\mathchar"218$}}
     \raise 2.0pt\hbox{$\mathchar"13C$}}}
\def\gta{\mathrel{\spose{\lower 3pt\hbox{$\mathchar"218$}}
     \raise 2.0pt\hbox{$\mathchar"13E$}}}

\begin{document}

\font\twelvei = cmmi10 scaled\magstep1 
       \font\teni = cmmi10 \font\seveni = cmmi7
\font\mbf = cmmib10 scaled\magstep1
       \font\mbfs = cmmib10 \font\mbfss = cmmib10 scaled 833
\font\msybf = cmbsy10 scaled\magstep1
       \font\msybfs = cmbsy10 \font\msybfss = cmbsy10 scaled 833
\textfont1 = \twelvei
       \scriptfont1 = \twelvei \scriptscriptfont1 = \teni
       \def\mit{\fam1 }
\textfont9 = \mbf
       \scriptfont9 = \mbfs \scriptscriptfont9 = \mbfss
       \def\bmit{\fam9 }
\textfont10 = \msybf
       \scriptfont10 = \msybfs \scriptscriptfont10 = \msybfss
       \def\bmsy{\fam10 }

\def\etal{{\it et al.~}}
\def\eg{{\it e.g.}}
\def\ie{{\it i.e.}}
\def\lsim{\raise0.3ex\hbox{$<$}\kern-0.75em{\lower0.65ex\hbox{$\sim$}}} 
\def\gsim{\raise0.3ex\hbox{$>$}\kern-0.75em{\lower0.65ex\hbox{$\sim$}}} 
\title{A SURVEY FOR INFALL MOTIONS TOWARD STARLESS CORES.\\
II. $\rm CS~(2-1)$ AND $\rm N_2H^+~(1-0)$ MAPPING OBSERVATIONS\footnote[1]
{To appear in Astrophysical Journal Supplement Series}}

\author{Chang Won Lee$^{1,2}$, Philip C. Myers$^{1}$, \& Mario Tafalla$^{1,3}$}
\vskip 0.2in
\affil{$^1$Harvard-Smithsonian Center for Astrophysics,}
\affil {60 Garden Street, MS 42, Cambridge, MA  02138, USA}

\vskip 0.2in
\affil{$^2$Taeduk Radio Astronomy Observatory, Korea Astronomy Observatory,}
\affil {36-1 Hwaam-dong, Yusung-gu, Taejon 305-348, Korea}
 
\vskip 0.2in
\affil{$^3$Observatorio Astron\'omico Nacional,}
\affil {Alfonso XII, 3, E-28014 Madrid, Spain}
\affil{E-mail: cwlee@cfa.harvard.edu, pmyers@cfa.harvard.edu, tafalla@oan.es}

\vskip 1in
\begin{abstract}
        We present the results of an extensive  mapping survey of 
`starless' cores in the optically thick line of CS(2-1)  and 
the optically thin lines of  
$\rm N_2H^+~(1-0)$ and $\rm C^{18}O~(1-0)$. The purpose of this
survey was to search for signatures of extended inward motions. 
A total of 53 targets were observed in the three lines with 
the FCRAO 14m telescope. Thirty three regions were mapped 
in both  CS and $\rm N_2H^+$, and thirty seven 
well-defined $\rm N_2H^+$ cores have been identified. 
The $\rm N_2H^+$ emission is generally compact
enough to find a peak,  
while the CS and $\rm C^{18}O$ emissions are more diffuse.
For each core, we have derived the normalized velocity
difference ($\delta V_{CS}$) between the thick CS and thin $\rm N_2H^+$ 
peak velocities. 
We define 10 `strong' and 9 `probable'  infall candidates, based on 
$\delta V_{CS}$ analysis and on the spectral shapes of CS lines. 

From our analysis of the blue-skewed CS spectra and the $\delta V_{CS}$ 
parameter, we find typical infall radii of $0.06-0.14$ pc.
Also, using a simple two layer radiative transfer model to fit 
the profiles, we derive one-dimensional infall speeds, half of whose values 
lie in the range of $\rm 0.05-0.09~km~s^{-1}$. These values 
are similar to those found in L1544 by Tafalla et al., and this result
confirms that infall speeds in starless cores are generally faster 
than expected from ambipolar diffusion in a strongly sub-critical core. 
In addition, the observed infall regions are too extended 
to be consistent with the  `inside-out' collapse model applied to a very 
low-mass star.
In the largest cores, the spatial extent of the CS spectra with
infall asymmetry is larger  than the extent of the $\rm N_2H^+$
core by a factor of $2-3$. All these results 
suggest that extended inward motions are a common feature in 
starless cores, and that they 
could represent a necessary stage in the condensation of a 
star-forming dense core.

\end{abstract}


\clearpage

\section{Introduction}

Inward gas motions are a key element of star formation.
Ambipolar diffusion or dissipation of turbulence can drive 
such motions during the process of core formation
(e.g., Shu et al. 1987, Ciolek \& Mouschovias 1995, Nakano 1998,
Myers \& Lazarian 1998), and gravitational collapse is the basis of
any model of stellar birth (e.g, Shu 1977). Despite the theoretical
expectation for a prevalence of inward motions in star-forming regions, 
little observational evidence for this type of kinematics still exists.
Only during the last decade, with its rapid development of 
mm-radio telescope instrumentation, have the first infall candidates
emerged (Zhou et al. 1993), and systematic work on infall motions
has been possible (Mardones et al. 1997, Gregersen et al. 1997).

Inward motions are inferred by observing 
the ``{\it infall asymmetry}'' in spectral lines: 
a combination of a double peak with a brighter blue component or 
a skewed single blue peak in an optically thick spectrum, 
and a Gaussian single peak in an optically thin line 
(Leung \& Brown 1977, Zhou 1995, Myers et al. 1996). 
As target sources, starless dense cores --regions 
which do not yet contain young stellar objects (YSOs)-- 
are excellent candidates, as they are expected to display
inward motions in the earliest stage of the star-forming
process, and therefore suffer from the least contamination by 
bipolar outflows. The recent observations of the starless core L1544
by Tafalla et al. (1998, hereafter T98) and Williams et al.
(1999) show that the infall asymmetry is spatially
too extended ($\sim 0.1$ pc)  to be consistent with the `inside-out' 
collapse model of Shu (1997), and that the large inward speed
(up to 0.1 km s$^{-1}$) is also too fast to result from ambipolar
diffusion in a strongly sub-critical core. This inconsistency of 
L1544  with the `standard' models casts many questions
on the physics of inward motions occurring in the very early stage
of star formation: Are the extended inward motions a
general phenomenon  that all cores should experience to form the star ?
If so, what would be the typical infall parameters of the collapsing
cores, e.g., the typical infall size, speed, and mass infall rate,
which may help us infer basic physics associated with inward motions ?   

To answer the above questions, a detailed study of infall motions in 
starless cores is necessary. 
Lee, Myers, \& Tafalla (1999) (hereafter LMT99)
have carried out the first such work by 
studying 220 starless cores with single pointing 
observations in the optically thick and thin tracers CS$(2-1)$ and
N$_2$H$^+~(1-0)$. These authors found that inward 
motions are statistically significant in starless cores.
A similar study of $\rm HCO^+~(3-2)$ emission 
found frequent infall asymmetry in 17 starless cores having sub-millimeter
continuum (Gregersen \& Evans 2000).
Single pointing observations, however, 
do not guarantee that each core with infall asymmetry is necessarily 
undergoing inward motions, because other kinematics such as differential
rotation and bipolar outflow can mimic that feature 
(Adelson \& Leung 1988, Zhou 1995). In order to clarify the real core
kinematics, mapping
surveys in both optically thick and thin spectral lines are therefore
necessary.

The goal of this paper is to complement the results of LMT99 with an
extensive mapping survey of dense starless cores in
optically thick CS (2--1) and thin $\rm N_2H^+ (1-0)$ 
[and $\rm C^{18}O(1-0)$] lines, so we can address
the questions suggested by the L1544 study.
This paper is structured as follows: 
in $\S$2 we explain details of the observations such as target selection,
observational equipment and methods, and data reduction. 
Detection statistics, mapping results, quantitative analysis of 
the infall size, speed, mass infall rate, 
and interesting individual cores are described, 
and implications of the observational results are discussed in $\S$3. 
In the last  section we summarize the main conclusions of this 
work.
 
\section{Observations}

Our mapping survey 
was  carried out in $\rm CS(2-1)$, $\rm N_2H^+~(1-0)$, and 
$\rm C^{18}O~(1-0)$ with the 14m telescope of the Five College 
Radio Astronomy Observatory (FCRAO)\footnote[1]
{FCRAO is supported in part by the National Science Foundation
under grant AST 94-20159, and is operated with permission of the Metropolitan
District Commission, Commonwealth of Massachusetts}. 
Targets were selected according to different criteria
(see last column of Table 1).
Thirty-four targets were selected based on the
single-pointing observations of LMT99. 
We observed 7 strong and 9 probable infall candidates defined by 
LMT99, as well as 18 targets showing strong CS and $\rm N_2H^+$ 
emission from LMT99. 
Twelve targets were selected from their $\rm NH_3$ detection 
by Benson \& Myers (1989, hereafter BM89), and seven targets
are from Lee \& Myers (1999, hereafter LM99). Overall, 
the starless cores in our target list are nearby 
(within a few hundred pc), have high extinction ($\rm A_v \sim 5^m$ 
based on the original Lynds selection, Lynds 1962), have optical sizes of 
$0.05\sim0.35$ pc (LM99), and typical line widths (FWHM of $\rm N_2H^+$) of  
$\rm 0.2 - 0.4~km~s^{-1}$ (Fig. 2 of this study).

In a quantitative study of infall motions using narrow line profiles, like
this one, it is important to use accurate frequencies 
for the different molecular lines (see LMT99).
Recent laboratory measurements (Gottlieb 2000, private
communication) have improved the precision of the frequency estimates 
for CS(2--1) ($97980.953 \pm 0.002$ MHz) and $\rm C^{18}O~(1-0)$
($109782.173\pm 0.002$ MHz), and in the rest of this paper 
we will use these new values. To obtain an equally accurate frequency for
the $\rm N_2H^+~(1-0)$ ``isolated'' ($\rm F_1F=01-12$) component, we have 
used new astronomical measurements of the extremely narrow lines toward
L1512, where taking as reference the above value of the 
$\rm C^{18}O~(1-0)$ frequency, we have derived an $\rm N_2H^+~(1-0)$  
frequency of $93176.258$ MHz (see appendix).
This new frequency set reduces the velocity shift 
between CS and $\rm N_2H^+$ in the LMT99 study 
by about 0.013 $\rm km~s^{-1}$. The effect of this change in the 
LMT99 conclusions is small, and is discussed in detail in the appendix.

Our extensive observations were made possible 
with the focal plane receiver arrays QUARRY and SEQUOIA at FCRAO. 
At the beginning of our survey (1997 January), the 
QUARRY system having 15 elements was used for a quick survey of 
about 42 targets in CS(2--1), $\rm N_2H^+$(1--0), and 
$\rm C^{18}O$(1--0). After that season, the new 
single sideband 16 element focal plane array receiver SEQUOIA
became available, and observations with this instrument 
were performed during 1998 March, April, and December, 1999 February and 
November, and 2000 January. The system temperature 
was typically  around 300 - 400 K for QUARRY and  140 - 180 K 
for the SEQUOIA, so most high quality data 
have been obtained with the SEQUOIA.
Thus, if data from both receivers are available, only the SEQUOIA data 
are presented. QUARRY has been designed to make a beam-sampled map of 
$200''\times 250''$ in a $50''$ grid with two pointings 
(which is  called one footprint map)
while SEQUOIA was used to make a beam-sampled map of 
$352''\times 352''$ in a $44''$ grid 
with four pointings (which is one footprint map). The telescope
beam size (FWHM) is approximately $52''$ at the CS(2--1) and
$\rm N_2H^+$(1--0) frequencies, and the main beam efficiency 
at these frequencies is about 0.56.

As back-ends, we used autocorrelation spectrometers with spectral 
channels of 10 and 20 KHz, achieving velocity resolutions of 
around $\rm 0.03 - 0.06~km~s^{-1}$. Such high resolutions 
are necessary for discerning the small amount of velocity shift 
between the optically thick and thin spectra. 
All observations were performed in frequency switching mode with a 
4 MHz offset for CS, and a 8 MHz offset for $\rm N_2H^+$ and 
$\rm C^{18}O$. The telescope pointing was checked and corrected
using 86 GHz SiO maser sources whenever the observing region was 
changed, and its accuracy was better than $10''$. 
The telescope focus was optimized prior to each pointing check.
In the QUARRY observation, all sources were observed in just one footprint
mode, and many extended sources were not fully covered.
During the SEQUOIA observations, we made several footprint
maps to cover the whole area of each core. 
Most of the spectral line data reduction was performed 
using the `CLASS' reduction software (Buisson et al. 1994).

\section{Results and Discussion}
\subsection{Detection Statistics}

Details of our observations of a total of 53 targets 
are summarized in Table 1.
Fifty out of 52 sources were detected in CS(2--1), 34 out of 45
were detected in $\rm N_2H^+~(1-0)$, and 28 out of 29
were detected in $\rm C^{18}O~(1-0)$.
Thirty-three sources were mapped in both the CS(2--1) and 
$\rm N_2H^+~(1-0)$ lines. It should be noted
that some sources consist of several smaller cores, 
while some sources were not fully mapped 
and their spatial extent is unknown.
Therefore, the number of fully mapped cores is different from
the number of observed or detected sources, and the central positions 
of the cores are sometimes different from the centers of the maps. 
We identify thirty-seven $\rm N_2H^+$ starless cores for which the spatial 
extent and the strongest positions are well defined, and we
list the coordinates, map sizes, velocities, line widths, and other
information of these cores in Table 2.
 
\subsection{Gas Distribution and Spectra}

We present a summary of our data in Fig. 1, where we
preferentially show sources with well defined 
$\rm N_2H^+~(1-0)$ distribution and interesting CS profiles.
For space reasons, we do not present sources for which 
$\rm N_2H^+~(1-0)$ was
not observed,  not detected, or partly mapped, and where the
CS profiles are noisy and show no asymmetric features.  
We have excluded L1445, for which $\rm N_2H^+$ is not detected although
its CS spectra show a self absorption feature while $\rm C^{18}O$
does not. The combination of CS and $\rm C^{18}O$ spectra in this source
suggests the presence of infall asymmetry, as in LMT99.    
We specially include $\rm N_2H^+$ maps for L1495 and  L1155C 
(Fig.  1-4 \& Fig. 1-20) to illustrate their multiple cores 
with extended $\rm N_2H^+$ distribution.

Most of the data  
are presented with a CS profile map 
superposed onto the $\rm N_2H^+$ integrated intensity distribution
on the left panel and with the averaged CS and $\rm N_2H^+$ spectra 
over the half maximum contour in the  $\rm N_2H^+$ intensity map 
on the right panel.
The dashed lines on the profiles are the systemic velocities derived
from Gaussian fits to the $\rm N_2H^+$ data  using the relative 
frequencies and intensities
of the 7 hyperfine components determined by Caselli et al. (1995).

As described in LMT99, there is a general trend of 
double peaks and asymmetric features in the CS spectra, while most 
of the $\rm N_2H^+$ and $\rm C^{18}O$ lines have a single peak and an 
almost Gaussian shape. Column 11 of Table 2 gives our estimates of 
the optical depth of the isolated ($\rm F_1F=01-12$)
hyperfine component, $<\tau_{N_2H^+}>$, defined as one-ninth of the 
total optical depth of the $1-0$ spectrum. This is determined by using the 
hyperfine structure (hfs) fitting routine in CLASS (Buisson et al. 1994), 
averaged over all spectra having S/N $>5$ in a map. We note
that the optical depth of the isolated $\rm N_2H^+$ component for most cores
is lower than 1.0, indicating that this component is optically thin.

To compare the spectra of the two thin tracers,
$\rm N_2H^+$ and $\rm C^{18}O$, we make
averages of the spectra with S/N $> 5$ for ten sources observed
with QUARRY ($\rm C^{18}O$ observations were conducted 
with QUARRY only). In this way we find that the
shapes of both lines are similar (Gaussian), while 
the $\rm C^{18}O$ line width (FWHM) is always larger
than that of $\rm N_2H^+$
by around $\rm 0.1-0.4~km~s^{-1}$. The difference between 
the line velocities of the two tracers is fairly  small (mean$\pm$ standard
deviation $\rm =0.007 \pm 0.078~km~s^{-1}$).

Spatially, the
$\rm N_2H^+$ emission is usually compact enough to define a core peak, 
while the CS and $\rm C^{18}O$ emissions are much more diffuse.
The greater extent of the CS and $\rm C^{18}O$ emissions could be due to 
their lower
effective critical density compared to that of $\rm N_2H^+$, while the
lower concentration of CS and $\rm C^{18}O$ toward the
core peak may be due to depletion of these
molecules or self-absorption in the case of CS, 
while $\rm N_2H^+$ experiences 
little depletion (e.g., Aikawa et al. 2001, Tafalla et al. 2001, 
in preparation) or self-absorption.
This greater extension of CS and $\rm C^{18}O$ compared to 
$\rm N_2H^+$ is similar to the pattern seen when comparing CS, 
$\rm C^{18}O$, and $\rm NH_3$ maps
in low-mass cores with and without YSOs (Myers et al. 1991)

Using the core size values presented in Table 2 (expressed as an
equivalent radius $\rm ({A / \pi})^{1/2}$, where 
A is area inside the $\rm N_2H^+$  half maximum contour),
we compare in Fig. 2 core sizes and line widths for the $\rm N_2H^+$
emission. The figure shows a slight indication of
a line width-size relation, which is similar to, but not as significant
as, that in Fig. 6 of Jijina, Myers \& Adams (1999)
using $\rm NH_3$ cores without IRAS sources. 
As seen in the figure, the typical sizes of the $\rm N_2H^+$ cores are 
between  $\sim 0.05$ pc and $\rm \sim 0.13$ pc, and 
the typical line widths of the cores are 
$\rm 0.2 - 0.4~km~s^{-1}$.

\subsection{Distribution of the Velocity Difference $ \delta V_{CS}$
Between the Optically Thick CS and the Thin $\rm N_2H^+$ Spectra}

The normalized velocity 
difference between an optically thick and a thin line
($\rm  \delta V$) has been 
found to be a sensitive measure of relative line shifts
indicative of inward motions
(e.g., Mardones et al. 1997, LMT99). For this work, 
we use the CS(2--1) and N$_2$H$^+$(1--0)  lines as thick 
and thin tracers, respectively, and define
$\rm  \delta V_{CS} = (V_{CS}-V_{N_2H^+})/\Delta V_{N_2H^+}$. 
For each spectrum in our 35 starless cores, we measure the velocity of 
the peak CS intensity
($V_{CS}$) from a Gaussian fit to the
brightest spectral component after masking the less bright component or 
the skewed part, and measure the N$_2$H$^+$(1--0) velocity 
$(V_{N_2H^+})$ from a 
hyperfine Gaussian fit to its seven components using the CLASS software.  
The FWHM of the N$_2$H$^+$ emission ($\Delta V_{N_2H^+}$) is also
derived from the hyperfine fit. 
To avoid ambiguity when choosing the bright CS component, we follow 
LMT99 and omit those CS spectra for which a Gaussian fit indicates 
a difference between the blue and red components smaller than the 
$1\sigma$ noise of the spectrum.

Table 2 presents our estimates of $\rm  \delta V_{CS}$ for each core,
together with estimates of the CS and N$_2$H$^+$ velocities derived 
using spectra with peak S/N larger than 5. Note that
the CS and N$_2$H$^+$ velocities, and the N$_2$H$^+$ line widths 
are averages over the core ($\pm$ standard error of the mean, or s.e.m): 
$\rm <V_{CS}>={\sum_i V_{CS}^i /N}$, 
$<V_{N_2H^+}>={\sum_i V_{N_2H^+}^i /N}$, and
$<\Delta V_{N_2H^+}>={\sum_i \Delta V_{N_2H^+}^i /N}$,
where N is the number of spectra with peak S/N larger than 5.
The parameter $<\delta V_{CS}>$ ($={\sum_i \delta V_{CS} /N}$) presented in 
the last column is 
the average ($\pm$ s.e.m) of all $\delta V_{CS}$s obtained in each core 
considering the CS and N$_2$H$^+$ spectra whose peak S/N is better than 5.
The number N of $\delta V_{CS}$ values and 
the radius R of the spatial extent
of the region having spectra with peak S/N better than 5
are given in columns 8 and 9.
R was estimated as half of the largest distance between core positions 
for which $\delta V_{CS}$s can be derived using spectra with S/N greater 
than 5. We have also estimated $\delta V_{CS}$ for 
the average CS and N$_2$H$^+$ spectra, but this parameter
is not significantly  different from $ <\delta V_{CS}>$
for most sources. Thus, its use instead of $ <\delta V_{CS}>$  
would not affect our conclusions.

The value of $<\delta V_{CS}>$  in each core is a useful indicator of the 
global overabundance of blue or red spectra, and suggests a
dominance of inward or outward motions. 
Fig. 3 plots the distribution of $ <\delta V_{CS}>$ for our sample, 
showing that most of our sources have a global 
overabundance of blue spectra characteristic of inward motions.
The excess of sources with overabundance of 
blue spectra in our sample, however, 
does not necessarily reflect the statistical infall excess in
starless cores found in the previous single pointing survey (LMT99),
because many sources  in our present sample have been selected with 
a bias toward infall candidates 
from previous observations. What the results in Fig. 3 do show is that
most infall candidates from the previous single 
pointing observations are now confirmed as infall candidates
when full maps are made, so the infall classification in LMT99 
does not arise from the observation of an special position, but
from a real property of the core as a whole.
  
\subsection{$ \delta V_{CS}$ Related Parameters and Classification of 
Starless Cores}

Here we introduce two $ \delta V_{CS}$ related parameters in order to
further measure the  distribution of asymmetric profiles
in a core and to classify cores in relation to inward motions.
One parameter is the blue excess $\rm E={(N_- - N_+) / N}$, 
where $\rm N_-$ is the number of positions with   
$\rm \delta V_{CS} \le -5~\sigma_{\delta V_{CS}}$, $\rm N_+$ is the number of
positions with $\rm \delta V_{CS} \ge 5~\sigma_{\delta V_{CS}}$, 
$\rm N$ is the total number of positions for which  $\delta V_{CS}$ 
was determined, and $\sigma_{\delta V_{CS}}$ is a propagation
error of $\rm \delta V_{CS}$. We have chosen
$n=5$ in $\rm n~\sigma_{\delta V_{CS}}$ for the limit value
to represent the degree of significance of the skewness in the CS profile
with respect to $\rm N_2H^+$ line. This is a compromise
between those which give too much noise (lower n) and too few cases to
allow analysis (higher n).
While $n=5$ is arbitrarily chosen, our
results are not sensitive to the exact choice  of $n$ for $n=4$ to 6.

The other parameter we introduce is 
a P value of a student t-test for the $ \delta V_{CS}$
distribution for each core, which is the probability  of
drawing our $ \delta V_{CS}$
distribution from a zero mean t-distribution. This P value is useful to
measure the statistical significance of the dominance of the blue or 
red asymmetry in the cores. A value very close to zero means 
that the distribution of the observed $ \delta V_{CS}$ is very far 
from the symmetric zero mean t-distribution.  The E and P values
for each core are listed in the last columns of Table 3. 

As Figures 4 and 5 illustrate, the E and P parameters
have a very close correlation with $< \delta V_{CS}>$.
Fig. 4 indicates that the blue excess increases as
$< \delta V_{CS}>$ becomes more negative, and Fig. 5 shows that
P approaches  zero as $ <\delta V_{CS}>$ deviates from 
zero (i.e., as a global overabundance of  blue or red asymmetry gets 
more significant).    
Combining these E and P parameters with additional spectral properties,
we now proceed to classify starless cores.
First, cores can be classified in two main groups:
cores with ``significant excess (SE)'' and  cores 
with ``no significant excess (NSE)''. Then, 
the ``SE'' cores would be divided into cores with 
``significant blue excess'' which would be further subdivided
into those with {\it double}  CS peaks and 
those with just {\it single} CS peak, and cores with 
``significant red excess''. 
The ``NSE'' cores might be 
subdivided  into a group where the spread in
$\delta V_{CS}$ is small, and there are few points 
where $\delta V_{CS}$ 
is significantly positive or negative, and into a group  where the spread in
$\delta V_{CS}$ is large, and there are several points with
significantly blue or red $\delta V_{CS}$.  

In Fig. 6 we display the number distributions of $ \delta V_{CS}$
for all cores according to the above classification, excluding
cores with a small number of $ \delta V_{CS}$ measurements ($<7$),
for which the parameters are not statistically significant,
and cores with noisy detection for which the number of positions
with measurements of S/N$>5$ is less than 3.
We now describe in more detail each of the groups.

Group 1 corresponds to cores with ``very significant blue excess''
and {\it double} CS peaks near the peak intensity position of $\rm N_2H^+$. 
These cores have high blue excess (E$\ga 0.35$)
and low P ($\la 0.1$), or significantly skewed $\delta V_{CS}$
distribution  to the blue (Fig. 6-a), corresponding to 
objects with overwhelming overabundance of blue asymmetric CS spectra.
Note that these criteria classify the well 
known infall candidate L1544 as a member of group 1, and we
will refer to the sources in this group as `strong infall candidates'.
This group consists of {\bf L1498, L1495A--S, L1544 
L1689B, L234E-S, L492, L694-2, L1155C-2}, and {\bf L1155C-1},  and possibly {\bf L183}, 
as discussed below.

Group 2 contains sources with  ``significant blue excess'' and
{\it single}-peaked CS spectra, which are blue-skewed,
near the peak intensity position of $\rm N_2H^+$. 
These cores have fairly high blue excess (E$\ga 0.1$) and low P ($\la 0.1$).
This group corresponds to `probable infall candidates' (Fig. 6-b),
and consists of
{\bf L1355, TMC2, TMC1, L1552, L1622A-2, L158}, and  {\bf L234E-C}, possibly {\bf L981} 
and {\bf L1197}, as discussed below. 

Group 3 contains sources with ``significant red excess'' having
high red excess (E $\la -0.15$) and $\delta V_{CS}$ distributions
significantly asymmetric
toward the red ($\rm P\approx 0.0$) (Fig. 6-c), but no blue
asymmetry in the CS spectra ($\rm N_-=0$). 
Three sources belong to this group: L1521F, L429-1, and CB246.

Group 4 corresponds to cores with ``no significant excess'' having 
small spread in $\delta V_{CS}$.
The sources in this group have little blue or red excess ($E\approx 0$), 
and have $\delta V_{CS}$ distributions 
with rather small rms (one standard deviation $\la 0.3$)  
(Fig. 6-d). Sources in this 
category are L1333, L1495B, L1400A, L1517B, L1622A-1, L1696A, and L234E-N.

Finally, group 5 contains cores with ``no significant excess'' and a
large spread in $\delta V_{CS}$.
These sources show little excess ($\rm E\approx 0$) either because 
they contain
a similar number of spectra of significantly blue or red $\delta V_{CS}$, or 
because they have few spectra of significantly blue or red $\delta V_{CS}$.
In any case, the distribution of $\delta V_{CS}$ has a
larger rms than that of group 4 (one standard deviation $\ga 0.4$) (Fig. 6-e).
Sources in this group are L1495A--N, L1507A, and L1512.

Although the above criteria seem to classify well the cores, it
is possible that they still miss some infall candidates.
L183, for example, has the largest angular area mapped in both
CS and $\rm N_2H^+$, and has a
different pattern of profiles depending on the region considered (Fig. 1-13).
Within the half maximum contour of $\rm N_2H^+$, L183
shows extended infall asymmetry, and the
statistics of its spectra are characteristic of a strong infall candidate
(E=0.6 and P=0.00). Toward the outer part of the core,
however, we find profiles with red asymmetry
(S-W of the core), skewed infall asymmetry, and 
symmetric lines, and when all the spectra are considered 
we obtain E=0.12 and P=0.62, which will not classify the core 
as an infall candidate. This different behavior of
the inner and outer regions is what one would expect if the
inner core has inward motions while the outer layers are static. For
this reason, we classify {\bf L183} as a strong infall candidate.

Other sources which may have a pattern similar to L183 and present
infall asymmetric profiles toward the inner core and
symmetric profiles outside are L1689B and L234E-S. Unfortunately,
the outer spectra in these sources are rather weak, and the infall 
pattern hard to discern. These sources deserve a more detailed study 
of their spectral properties.

As mentioned before, 
we have not included in our classification sources with a small number of 
$\delta V_{CS}$ measurements ($< 7$) and sources with noisy detections. 
Sources excluded in this way are
CB23, L134A, L981-1, L1063, and L1197. Note that 
{\bf L981-1} and {\bf L1197} show typical infall profiles in their
compact $\rm N_2H^+$ cores, so they are likely  to be considered 
probable infall candidates when more sensitive observations are carried out.

\subsection{Infall Candidates from Mapping Observations and from 
Single Pointing Observations}

In our survey we have mapped 15 infall candidates suggested by LMT99.
We now check whether these infall candidates from single-pointing 
observations continue being infall candidates after a complete
mapping observation. 

According to the above core classification, 11 of the 15 LMT99 sources 
are now classified as strong or probable infall candidates.
These sources are L1355, L1498, L183, L158, L694-2, L1155C-1,
TMC2, L1622A-2, L1689B, L234E-S, and  L492 
(the source name L234E-1 used in the previous
single pointing observations has been changed to L234E-S 
in this survey).

L1521F was considered a strong infall candidate in LMT99 
because of its blue asymmetry in the CS profile, but 
in our mapping survey it shows a 
reverse asymmetry -- {\it red} asymmetry --
which is also extended. As the Haystack telescope beam used in 
LMT99 is about one half of the FCRAO beam, the infall asymmetry
in the smaller beam and the red asymmetry in the larger beam 
suggest that in this core the direction of the gas motions may change
toward the center. More observations are needed to 
understand this change.

CB23 was classified as a probable infall candidate by LMT99,
and our mapping survey finds a slight asymmetric CS profile 
at the center. However, all other profiles have a Gaussian shape,
making the E value of this source equal to zero. The small 
number of $\delta V_{CS}$ measurements in this core (4) leaves
it unclassified.

Another source which remains unclassified in this study, despite 
being a probable infall candidate in LMT99 is L1445, as it
was not detected in $\rm N_2H^+$ by our FCRAO observations. 
The LMT99 infall candidates L1524-4 and B18-3 have not
been mapped with the FCRAO telescope, so their infall status cannot
be tested. Finally, new infall candidates not considered as such by LMT99
include TMC1, L1552, and L1155C-2, and all show extended 
infall asymmetry in their profile maps.

In summary, most infall candidates from the single 
pointing observations remain bona fide candidates, although
a minority of sources show spectral profiles different from 
those inferred using the single pointing data.  

\subsection{Infall Radius, Speed, and Mass Infall Rate of 
Starless Infall Candidates}

We now estimate the infall radius, speed, and mass infall rate for our
infall candidates in groups 1 and 2, as well as for L183, L981-1, and L1197.
We consider two cases depending on the 
choice of spectra used to calculate the infall parameters. 
Case 1 is called the `weak' case (indicated by `W'), and considers all blue
asymmetric spectra satisfying  
$\rm \delta V_{CS} \le -5~\sigma_{\delta V_{CS}}$.
Case 2 is referred as the `strong' case (indicated by `S'), and
considers only the CS spectra with double-peaked infall asymmetry satisfying 
$\rm \delta V_{CS} \le -5~\sigma_{\delta V_{CS}}$.

\subsubsection{\bf Infall Radius}

We estimate the infall radius of each core from the spatial distribution 
of $\rm \delta V_{CS}$ by measuring the regions where $\rm \delta V_{CS}$ is 
significantly negative. We define the infall radius as   
one half the largest distance between positions
showing infall asymmetry of 
$\delta  V_{CS} \le -5~\sigma_{\delta  V_{CS}}$ in their spectral profiles. 
For L158, L234E-C, and L1197, which have only one position with
$\delta  V_{CS} \le -5~\sigma_{\delta  V_{CS}}$,
we take as infall radius one half the linear size of the telescope beam 
FWHM ($\sim 26''$). 
Table 4 gives our estimates of the infall radii of the cores 
for the `weak' and `strong' cases.

Our method of deriving infall radii from the $\rm \delta V_{CS}$ distribution
is limited by the spatial extent of the $\rm N_2H^+$ emission.
In some cores (e.g., L1689B and L158), the region with 
blue-skewed CS spectra is more extended than the region with detected 
$\rm N_2H^+$, so a method based on the $\rm \delta  V_{CS}$ 
distribution gives infall radii smaller than one based on 
the spatial extent of the blue-skewed CS profiles. 
Thus, as an alternative, 
we also measure infall radii by using the spatial extent of the blue-skewed 
CS profiles, taking as infall radius  one half the largest separation 
between  positions with skewed CS spectra.  

We present the estimated infall radii for all three 
cases in Table 4, and show their number distribution in Fig. 7.
The statistics show that infall radii are typically  about
0.06 -- 0.10 pc for the `weak' and `strong' $\rm \delta V_{CS}$ cases,
and around 0.06 -- 0.14 pc for the estimates based on the extent of 
the blue-skewed CS profiles.

\subsubsection{\bf Infall Speed}

We estimate infall speeds of the `weak' and `strong' cases  for our infall
candidates by applying a simple two layer model
similar to that of Myers et al. (1996). For this, we
average all blue asymmetric spectra satisfying the `weak' and `strong' cases,
respectively. 

Our two-layer model consists of a cool (2.7 K) absorbing front screen 
moving away from us and a warm emitting rear layer approaching us 
with same speed. The emerging brightness distribution in terms
of the line brightness temperature $\rm \Delta T_B$ is given
by (see Myers et al. 1996):
$$
\Delta T_B=J(T_r)(1-e^{-\tau_r})e^{-\tau_f}
+J(T_b)(e^{-\tau_f-\tau_r}-e^{-\tau_f}),
$$
where $J(T)={T_0 / [exp(T_0/T)-1]}$, $T_0=h\nu/k$, $\nu$ is the 
frequency,
$h$ is Planck's constant, $k$ is Boltzmann's constant, $T_b$ is the cosmic
background temperature (2.7 K), $T_r$ is the excitation temperature of
the rear layer, and $\tau_f$ and $\tau_r$ are the optical 
depths of the front and rear layers, respectively. 
The optical depths are given by 
$\tau_f=\tau_0 \exp[-(v-V_f)^2/2\sigma^2]$, and 
$\tau_r=\tau_0 \exp[-(v+V_r)^2/2\sigma^2]$, where $\tau_0$ is the peak 
optical depth,  $V_f$ and $V_r$  are
the systemic velocities of the front and the rear layers, respectively,  
and $\sigma$ is the velocity dispersion of the gas in both layers.

Given the above equation, our fitting procedure has 
5 free parameters [$\tau_0$, $V_f=V_r(=V_{in,z}/2)$, $T_r$,
$\sigma$ ($=\Delta V_{FWHM}/2.355$), and $V_{LSR}$]. 
Usually, $\sigma$ and $V_{LSR}$  are very well constrained by the width and
centroid velocity
of the observed spectra. The rest of the free parameters
depend on different features of the line profile.
The peak optical depth $\tau_0$  is sensitive to the depth of 
the self-absorption dip, the infall speed $V_{in,z}$ primarily affects  
the ratio of blue to red peaks of a self-absorbed spectrum,
and the excitation temperature of the rear layer $\rm T_r$ controls 
the overall intensity scale of the spectrum.  As these parameters affect
different aspects of the line profile, they are usually well 
constrained, and tests repeating the fitting procedure return
very similar values of these parameters.

The results of our modeling are presented in Fig. 8, where
model spectra (thick solid lines) are shown superposed to the 
observed spectra (thin solid histograms). For space reasons, we
show
results of the weak case for the cores without strong case,
and of the strong case only for cores with both cases. 
(L694-2 is the exception).
Note that spectra for the two cases are only slightly different,
and as it can be seen, the fit is quite good in most cases.
The numerical results of the non-linear least squares fitting
procedure are listed in Table 4.

The most important parameter derived from our model
fitting is $\rm V_{in,z}$, so we have estimated its uncertainty
with a semi-empirical procedure.
We have taken a best-fit spectrum for L694-2 (strong case), added random
noise with the same rms as the observed noise, and made a
new fit. By repeating this procedure 100 times, 
we have derived a 1 sigma uncertainty 0.001 $\rm km~s^{-1}$,
which is much smaller than the estimated $\rm V_{in,z}$
(0.086 $\rm km~s^{-1}$). Although the uncertainty 
may be slightly larger for 
cores where the fit is not as good as in L694-2 (e.g., L1155C-2),
the above estimate suggests that our $\rm V_{in,z}$ values 
are not dominated by noise.

To summarize the $\rm V_{in,z}$ estimates, we present in Fig. 9 a
histogram of this parameter, showing that most infalling starless cores 
have $\rm V_{in,z}$ of $\rm 0.05 - 0.09~km~s^{-1}$ for both the `weak' 
and the `strong' case. It should be noted, however, that
the infall speed derived here is the 
line-of-sight velocity, which is smaller than the three dimensional (3-D)
infall speed ($V_{in}$) by a likely factor of about $1-2$. 
Comparing our estimates with those of LMT99, we find a reasonable agreement,
despite the fact that the LMT99 values were simply
derived by taking half of the velocity dispersion
of the $\rm N_2H^+$ line obtained from a single pointing observation. 

\subsubsection{\bf Mass Infall Rate}

The mass infall rate determines how fast matter is condensing 
through the surface
of the infall radius ($\rm R_{in}$). We estimate this rate 
from  $4\pi R_{in}^2\rho V_{in}$,  assuming that the core has uniform 
density $\rho$ and is spherically infalling with a speed $\rm V_{in}$. 
Our two-layer model does not constrain well 
the density of the gas causing the absorption dip or shoulder in the 
CS profiles, so we approximate it by
$\rho$ ($={{15 \sigma^2}/{4\pi G R_{in}}}$), which is the density
of a core in virial equilibrium with radius ($\rm R_{in}$) and velocity 
dispersion ($\sigma$) (see column 11 of Table 4).
Here the velocity dispersion is derived from the two layer model fit.
In this way, gas densities are between about 
$\rm 3.0 \times 10^3$ and  $\rm 4.1 \times 10^5~cm^{-3}$, 
with a typical value of a few $\rm 10^4~cm^{-3}$.

The resulting mass infall rates for the strong and weak cases are 
between a few  $\rm 10^{-6} \sim 10^{-5}~\msun ~yr^{-1}$.
Individual estimates for  each source are given 
in the last column of Table 5. All these estimates are 
derived using the projected infall speed ($\rm V_{in,z}$), so
the true mass infall rates, derived from the  3-D infall speeds,
will be probably larger by a factor of $1-2$.

\subsection{Infall Radius and Core Size}

A comparison of the size of the infall region with
the core size is presented in Fig. 10 as a plot of
the infall radius versus the radius of
the $\rm N_2H^+$ emission (listed in the 4th column of Table 2).
Three cases are presented in the figure according to the method used in
deriving the infall radius: (a) weak case, (b) strong case, and (c)
case using the extent of the blue-skewed CS profiles. 

The errors for both radii are difficult to specify for each source.
The uncertainty of distance would be a major cause for the errors,
however, such uncertainty would equally change both radii
in the same way and so the tendency shown in Fig. 10 would not be
affected due to the uncertainty of distance.
The uncertainties of both radii which we can approximately give are
the values obtained
from repeating measurements, which are typically small (about between
$0.01 - 0.02$ pc) compared with the estimates of the radii.

Dashed lines in the figure correspond to core sizes
equal to infall sizes. As seen in the figure,
in strong case (b) the data are along the dashed line, i.e.,
larger cores tend to have equivalently larger infall zones.
This is because the infall size in the strong case is strongly limited with
an extent of distribution of CS profiles with double peaks
which is localized around the central region of the $\rm N_2H^+$ core.

In (a) and (c) cases, however, the slope of the data is even steeper
than that of the dashed line,
i.e., large cores tend to have a factor of about $2 - 3$
larger infalling zones than their extents,
implying that the core tends to have wider infalling region as the core is 
becoming bigger, or vice versa.
This result suggests
that infall motions that we are seeing  are likely involved in the building-up
of the dense core.

\subsection{Possible Interpretation of Extended Inward Motions}

Taking as a whole the results in Table 4, we can see that
the main features of the inward motions 
in our infall candidates are very similar 
to those found in L1544 by T98 and Williams et al. (1999).
This indicates that L1544 is not exceptional in having extended
infall asymmetry, and we interpret this as a suggestion 
that the extended infall phenomenon
may be a necessary stage in core evolution 

To investigate the cause of the extended inward motions in starless
cores we explore different alternatives.
First, we believe that these motions do not represent 
directly the process of forming a star, but of building up a dense core. 
Our cores have been selected to be starless in the sense that 
there is no associated IRAS point source (LM99), and so
far no other evidence for embedded stellar objects in these 
cores has been found.
Instead, there are supporting clues that our targets are starless,
as the fact that the spectral lines from all sources are 
very narrow ($\rm N_2H^+$ FWHM of about $\rm 0.2-0.4~km~s^{-1}$), 
and the fact that
no source shows wide wing components in its
profiles suggestive of the presence of outflow motions. 
This lack of evidence for star formation makes the extended 
inward motions (at least 0.06 pc in radius) incompatible with the
``inside-out'' collapse model of Shu (1977), because the
large infall sizes imply large times since collapse started, and
require the presence of detectable
YSOs at the core centers, something we can easily rule out.

Alternative mechanisms for producing extended inward motions involve 
the process of ambipolar diffusion, or the dissipation of
turbulent motions.  Infall speeds, however, are usually over 
$\rm \sim 0.05~km~s^{-1}$, which is too fast to result 
from ambipolar diffusion in a strongly sub-critical core 
(e.g., `standard' $\rm B_{UB}$ model of Ciolek \& Mouschovias 1995
predicting $\rm \sim 0.02~km~s^{-1}$ at 0.05 pc). Still,
the possibility of ambipolar diffusion can not be ruled out.
For example, recent models for super-critical cores 
are found to produce infall speeds closer to $\rm \sim 0.1~km~s^{-1}$ at
0.05 pc (e.g., Ciolek \& Basu 2000),
and models of slightly sub-critical cores predict
inward speeds intermediate between those of strongly sub-critical 
ambipolar diffusion and of dynamical motions 
(Zweibel 1998; Indebetouw \& Zweibel 2000).
Furthermore, a spherical core model which includes the effects of dust 
grains on field-gas coupling predicts extended inward motions
of $\rm \sim 0.1~km~s^{-1}$ (Li 1999).

Alternatively,  turbulent dissipation models 
(e.g., Myers \& Lazarian 1998, Myers \& Zweibel 2001) also predict
extended inward motions, substantially faster ($\rm \sim 0.1~km~s^{-1}$) 
than those from ambipolar distribution of a strongly sub-critical core. 

Thus, super-critical, slightly sub-critical ambipolar diffusion, or 
turbulent dissipation 
may explain the presence of the extended inward motions
we find, although more detailed models are still needed to settle
this problem.

\subsection{Individual Sources}
In this section we briefly  describe the kinematical characteristics and
interesting features of some sources.
We especially focus on those sources for which 
high S/N data were obtained. 
Detailed descriptions on the infall structure of some infall 
candidates will be presented in a future study.

{\bf L1498}-- 
Our integrated intensity map of $\rm N_2H^+$ for L1498 
shows that it has a well-defined gas distribution elongated
NW to SE with a position angle of about $-50^\circ$ (Fig. 1-3). 
This source is well studied by several authors
(e.g., Lemme et al. 1995, Kuiper, Langer, \& Velusamy 1996,
Wolkovitch et al 1997, and Gregersen \& Evans 2000). 
L1498 has been suggested to be a classic example of a 
pre-protosteller core that may be on the verge of rapid collapse
to form a protostar from CCS, $\rm NH_3$, $\rm C_3H_2$,
$\rm HC_7N$, $\rm C^{18}O$, and $\rm ^{13} CO$ observations
(Kuiper, Langer, \& Velusamy 1996).
Lemme et al. (1995) have first found the extended distribution of 
CS (3-2) profiles with double peaks of the blue peak 
brighter than the red peak. They interpreted this
might be either from absorption by a foreground low density cloud
or from presence of two layers of gas with slightly different 
line-of-sight velocities, with a double peaked 
feature of the optically thin tracer, $\rm C^{34}S$ which was poorly detected. 
Our observation rules out the second possibility because the
optically thin tracer $\rm N_2H^+$, detected with high S/N,
shows clearly single peaked spectra, as seen in Fig. 1-3.

Infall asymmetry in the CS profiles  is prominent and  spatially extended. 
Interestingly, CS spectra of skewed single peak and double 
peaks are spatially mixed.
The CS profile map shows infall profiles
of double peaks near the peak $\rm N_2H^+$ intensity and 
single skewed peak outside. 
Average profiles over the HM contour area of $\rm N_2H^+$ intensity
show typical infall asymmetry, suggesting that inward motions are
globally dominant over the core. 
Such an indication is also seen in the distribution of $\delta V_{CS}$,
which is overwhelmingly skewed to the blue (Fig. 6-a).
This source is suggested to be a `strong' infall candidate.
Possible infall radius and speed are $0.05- 0.08$ pc 
and $\rm 0.03-0.08~km~s^{-1}$, and the derived mass
infall rate is
$\rm 0.09-0.13 \times 10^{-5}~\msun ~yr^{-1}$.

{\bf L1495}--  The $\rm N_2H^+$ integrated intensity map in Fig. 1-4 shows 
that this core has three distinct $\rm N_2H^+$ condensations:
L1495A--N, S, and L1495--B.
{\bf L1495A--N} is elongated north-south (Fig. 1-5), and
its extended infall asymmetry of CS profiles reverses 
to extended red asymmetry with respect to a NE--SW axis which
crosses the center of the core
(drawn with solid line in Fig. 1-5), suggesting the presence of 
differential rotation. 

Fig 1-5 also shows the $\bf L1495A-S$ core, whose 
$\rm N_2H^+$ emission is elongated approximately E-W.
The CS spectra toward L1495A-S consistently show 
infall asymmetry, both in the form of skewed spectra and 
double-peaked profiles, like L1498. 
The infall radius inferred from the $\delta V_{CS}$ distribution is about 
0.09 pc, and the value from the skewed CS profiles  
is slightly larger ($0.12$ pc). The infall speed is estimated in
$0.07- 0.08 \rm~km~s^{-1}$, and the mass infall is 
$\rm 1.0 - 1.1 \times 10^{-5}~\msun ~yr^{-1}$.

The third L1495 core, shown in Fig. 1-5, is L1495B, and lies $\sim 8'$ west
of L1495A--S. This core is weaker and smaller than the other 
L1495 cores.  In contrast with those cores, L1495B has 
few significantly asymmetric CS  spectra. The symmetric $\delta V_{CS}$
distribution (Fig. 6-d), the mean $\delta V_{CS}$ value of 
$0.04\pm 0.13$, and high P from the t-test
(Table 2 and 3) are consistent with no evidence for 
inward motions in this core.

{\bf L1521F}-- Observations of this source show that features of 
the spectral asymmetry
can be dependent on the observing tracer or the observing spatial resolution.
L1521F has been  classified as an infall candidate by LMT99 based on their 
high angular resolution CS(2--1) observations, and by
Onishi et al. (1999) from their $\rm HCO^+(3-2)~\&~(4-3)$ data
(note that these authors named this core as MC27).
Our map, however, indicates an extended red asymmetry 
over the core (Fig. 1-6). Further study of this object in different
tracers is needed to clarify its kinematics.

{\bf TMC2}-- The $\rm N_2H^+$ emission  of this core is extended
(about $6'\times 6'$) both N-S and E-W (Fig. 1-7).
Most CS spectra show skewed blue asymmetry, and the
Gaussian fit velocities of $\rm N_2H^+$ 
lie on the skewed red part or the red shoulder of the CS spectra, 
suggesting possible extended inward motions. The infall radius of 
this core is somewhat large ($\rm 0.13 \sim 0.14$ pc) and 
the infall speed is the largest in our sample ($\rm \sim 0.17~km~s^{-1}$),
making the mass infall rate also the largest one
($\rm 4.0 \times 10^{-5}~\msun ~yr^{-1}$). 

{\bf TMC1}-- TMC1 has been the best target in Taurus for an extensive 
study of the physics of star formation and the chemistry 
(e.g., Hirahara et al. 1992, Langer et al. 1995, and Pratap et al. 1997).
It is known to be very extended, about $12'$ by over $35'$ 
in $\rm C^{18}O~(2-1)$ (Langer et al. 1995). 
Our map shows only a part around the center of TMC1.
The distribution of $\rm N_2H^+$ is extended  NW - SE (Fig. 1-8).
Some CS spectra show blue-skewed asymmetric shapes, while others
are closer to Gaussian.
However, the Gaussian fit velocities of the $\rm N_2H^+$ spectra
are located somewhere toward the red of the CS spectra, 
regardless of the CS line shape.
Moreover, careful inspection of the average CS spectrum on the 
right panel in Fig. 1-8
indicates that TMC1 may have one or two 
different velocity components shown as two wings to each side
of the strong peak component.
A two-layer model fit to the strong peak component gives
a possible infall speed of $\rm \sim 0.05~km~s^{-1}$.
The estimated infall radius is at least $\sim 0.16$ pc, and
the mass infall rate is about
$\rm 0.6 \times 10^{-5}~\msun ~yr^{-1}$.

{\bf L1512}-- The distribution of $\rm N_2H^+$ emission from this
core is rather compact and round. The line width of this 
source is the narrowest ($\rm <\Delta V_{N_2H^+}> \approx 0.19~km~s^{-1}$)
among our sources.
The $\rm HCO^+~(3-2)$ spectrum which Gregersen \& Evans (2000) have 
obtained toward this source shows no asymmetric feature. This may be partially 
because the spectral resolution ($\rm 0.12-0.15~km^{-1}$)
that they used was not sufficient to resolve it.
On the other hand, our observation shows clear asymmetric features in spectra
which are also changing over the core.
As in  L1495A--N, the sense
of the asymmetry in the CS spectra changes
from blue to red with respect to an 
SW -- NW axis (indicated with solid line in Fig. 1-9),
again suggesting differential rotation.

{\bf L1622A}-- This core has two $\rm N_2H^+$ condensations forming
a NE-SW filament (Fig. 1-12). We refer to the 
weaker NE condensation as {\bf L1622A-1}, and to the brighter SW condensation
as {\bf L1622A-2}. Both condensations  have similar systemic $\rm V_{LSR}$ 
($\rm \approx 0.2- 0.3~km~s^{-1}$), but their
$\delta V_{CS}$ distributions are different, as
L1622A-2 has more blue asymmetric CS spectra
(Fig. 6-b) than L16222A-1 (Fig. 6-e). 
Because of its blue asymmetric spectra, L1622A-2 is 
selected as a probable infall candidate. The estimated
infall radius, speed, and mass infall rate for this core are
$0.24- 0.26$ pc, $\rm \sim 0.09~km~s^{-1}$, and
$\rm 1.2 \times 10^{-5}~\msun ~yr^{-1}$, respectively.

{\bf L183}--  This core is one of the largest
in our sample, and mapping its total extent required covering an area of
$\sim 11' \times 18'$. As Fig. 1-13 shows, the $\rm N_2H^+$ emission
lies mostly along the N-S direction. In the BM89 
catalog, the brightest position was referred to as L183B and 
the weaker position as L183. Here, we treat the system as a single core
(L183) because there is no clear boundary between the peaks 
in the $\rm N_2H^+$ emission.

The spatial distribution of self-absorbed CS spectra is rather 
complicated as described in $\S$3.4.
The number distribution
of $\delta V_{CS}$ of all spectra of this source is rather close
to symmetric. 
However, this source was classified as a strong infall candidate because
the inner region (within the HM) of the $\rm N_2H^+$ core shows high
blue excess and a significantly asymmetric $\rm \delta V_{CS}$ distribution
to the blue, while the outer region does not (Fig. 1-13 and Fig. 6).
A more detailed study is needed to understand the
kinematics of L183.

{\bf L1689B}-- L1689B is known to be a small core 
(FWHM mass of $\sim 0.6 M_{\odot}$) located at the east edge of L1689 complex 
of Ophiuchus star forming region (Loren 1989, Andr\'e et al. 1996).
Our $\rm N_2H^+$ map shows a structure with a very compact 
center and a diffuse envelope elongated NW -- SE (Fig. 1-14).

Several molecular line observations toward this core
[$\rm CS~(2-1)$; LMT99, $\rm H_2CO~(2_{12}-1_{11}$; Bacmann et al. 2000,
$\rm HCO^+~(3-2)$; Gregersen \& Evans 2000]
have revealed a typical infall asymmetry in their spectra.
It has been suggested that this core is just 
entering the phase of dynamical contraction 
following the formation of a super-critical 
core, as revealed from 1.3mm continuum, and ISOCAM absorption studies
by Andr\'e et al. (1996) and Bacmann et al. (2000).

Our spectral line observations strongly support the above suggestion.
The CS spectra present extended infall asymmetry, in the form of
double peaks, blue peak with a red shoulder,
and blue-skewed peaks, implying extended inward motions over the source.  
The infall radius, speed, and mass infall rate 
are $0.08-0.18$ pc, $\rm \sim  0.05~km~s^{-1}$, 
and $\rm \sim 0.5 \times 10^{-5}~\msun ~yr^{-1}$, respectively. 

{\bf L234E}-- This region has three well defined $\rm N_2H^+$ cores:
{\bf L234E-N, C,} and {\bf S} (Fig. 1-16). {\bf L234E-S}, the
most southern one, 
has the largest negative $<\delta V_{CS}>$ ($-0.63\pm 0.14$), 
a blue asymmetric $\delta V_{CS}$ distribution, and
a low P value ($\approx 0$), which make it a strong infall candidate. 
Most CS spectra present double peaks or a blue peak with 
a red shoulder. We estimate the core
infall radius, speed, and mass infall rate are
$0.06 - 0.12$ pc, $\rm 0.05 ~km~s^{-1}$,
and  $\rm 0.4 - 0.6 \times 10^{-5}~\msun ~yr^{-1}$, respectively.

{\bf L234E-C} also show infall asymmetry 
at some positions, but not as many as L234E-S. 
This source has one position with
$\rm \delta V_{CS} \le -5~\sigma_{\delta V_{CS}}$, so
its infall radius is estimated from one half the telescope FWHM,
corresponding to $0.02$ pc. The infall speed and mass infall rate are
estimated in 0.05 $\rm km~s^{-1}$ and
$\rm \sim 0.33 \times 10^{-5}~\msun ~yr^{-1}$.

{\bf L234E--N} has no position with 
$\rm \delta V_{CS} \le -5~\sigma_{\delta V_{CS}}$,
although its $\delta V_{CS}$ distribution is skewed to the blue.
We classify this as a member of Group 4.

{\bf L429-1}-- The $\rm N_2H^+$ peak emission in this core lies 
to the NW of the CS emission (Fig. 1-18), some
$\sim 88''$ away
from the map center position which was defined by the 
optical minimum in the Digitized Sky Survey image (LM99).
Most CS profiles show a very strong self-absorption feature 
with brighter red peak, as found in L1521F.

{\bf L694-2}-- This is one of our strong infall candidates. 
The CS profile map shows double peaked infall profiles near 
the $\rm N_2H^+$ peak region, 
which change to blue-skewed and symmetric at larger separation
(Fig. 1-19).
The distance to L694-2 is not yet known, although we assume 
it is the same as that to B335 (250 pc),
because L694-2 lies within a few degrees of B335 and has 
similar LSR velocity. 
Infall radius, speed, and mass infall rate are
$0.09 - 0.13$ pc, $\rm 0.05 - 0.07 ~km~s^{-1}$, and
$\rm 0.4 - 0.6 \times 10^{-5}~\msun ~yr^{-1}$, respectively.

{\bf L1155C}-- L1155C is a dense molecular 
condensation within a dark cloud L1158 in
Cepheus. Harjunp\"a\"a et al. (1991) have observed this source in $\rm CO$,
$\rm ^{13}CO$, $\rm C^{18}O$, $\rm HCO^+$, and $\rm NH_3$, and have
found two cores with slightly different velocities. These authors
suggest that the cores might be a bound pair. 
From a test of Jeans instability, one core (L1155C-2) has been 
suggested to be probably collapsing. Note that we
we have named the bright core
as L1155C-1, and the weak one as L1155C-2, following LMT99. 
Harjunp\"a\"a et al., however, have referred to them 
in the opposite way.

Our observations (Fig. 1-20) also shows two distinct cores  
with slightly different systemic velocities,
L1155C-1 ($\rm V_{LSR}\approx 2.7~km~s^{-1}$) and 
L1155C-2 ($\rm V_{LSR}\approx 1.4~km~s^{-1}$).
 
The CS spectra on the right panels in Fig. 1-21 and 22
show additional velocity components superposed along the line of sight. 
Both cores show a significant number of infall profiles,
and are classified as strong infall candidates.
Their $\delta V_{CS}$ distributions shown in Fig. 6-a clearly indicate
overabundance of blue asymmetric CS spectra (both
double peaks and single peaks with infall asymmetry).
{\bf L1155C-2} is elongated  N-S and twice as large as 
L1155C-1 (Fig. 1-21). It has a large velocity gradient 
along the major axis ($\rm \sim 0.19~km~s^{-1}\  arc min^{-1}$),
which can confuse the interpretation of the simple average CS spectrum.
For this reason, 
we present in the right panel of Fig. 1-21
the spectra averaged within  the HM contour of the $\rm N_2H^+$ 
emission  after coinciding the systemic velocities of the spectra 
with that at peak position [$\rm (\Delta \alpha, \Delta
\delta)=(-88'',-132'')$].
The estimated infall radius is the largest in our sample
($0.12 - 0.45$ pc), and the 
infall speeds and mass infall rates are 
$\rm 0.07- 0.10~km~s^{-1}$ and 
$\rm 0.4 - 1.2 \times 10^{-5}~\msun ~yr^{-1}$, respectively.
{\bf L1155C-1} is located NE of L1155C-2 and has 
a smaller, more compact $\rm N_2H^+$ distribution 
(Fig. 1-22). There is no significant velocity gradient, 
so the profiles shown in
Fig. 1-22 are averages over the HM contour of the 
$\rm N_2H^+$ emission. 
The infall radius, speed, and mass infall rate for this core are 
$0.15 - 0.25$ pc, $\rm 0.09 - 0.1~km~s^{-1}$, and 
$\rm 2.3 \sim 2.5 \times 10^{-5}~\msun ~yr^{-1}$, respectively.

\section{Summary}
        We have carried out an extensive  
mapping survey of starless cores in optically thick [$\rm CS(2-1)$]  and thin 
[$\rm N_2H^+~(1-0)$ and $\rm C^{18}O~(1-0)$] lines searching
for infall candidates and studying the
general features of inward motions in the 
earliest stage of star formation. 
The  FCRAO 14-m telescope equipped with the focal plane receiver array
systems QUARRY and SEQUOIA
was used for these observations. Fifty out of
52 cores were detected in CS(2--1), 34 out of 45 were detected in 
$\rm N_2H^+~(1-0)$, and 28 out of 29 were detected in $\rm C^{18}O~(1-0)$.    
A total of 33 regions with strong emission have been mapped 
in both  CS and $\rm N_2H^+$, identifying 37 well-defined 
$\rm N_2H^+$ starless cores.
The $\rm N_2H^+$ emission is usually very compact
and shows a well defined peak, 
while the CS and $\rm C^{18}O$ emissions are much more 
diffuse. CS spectra in a large number of starless cores show
profiles with infall asymmetry,
while most of the $\rm N_2H^+$ spectra  
show a single Gaussian profile in their isolated component. 
The typical sizes and line widths (FWHM) of the $\rm N_2H^+$ cores are 
found to be between  $\sim 0.05$ pc and $\rm \sim 0.13$ pc, and 
 $\rm 0.2 - 0.4~km~s^{-1}$, respectively.

We have quantified the kinematical properties of each core
by using the normalized velocity
difference $\delta V_{CS}$ between the thick CS and thin $\rm N_2H^+$ peak 
velocities. We have also defined $\delta V_{CS}$-related parameters,  
such as the blue excess E and the t-test probability value P. Using these 
parameters and the shapes of the CS lines,
we have classified the starless cores into five distinctive groups:

(1) cores with ``very significant blue excess'' having 
overabundance of blue asymmetric CS spectra
with {\it double} peaks--{\bf L1498, L1495A--S, L1544, L1689B, L234E-S,
L492, L694-2, L1155C-2, and  L1155C-1,} and possibly {\bf L183}.
 
(2) cores with ``significant blue excess'' having 
overabundance of blue asymmetric CS spectra
with a  {\it single} peak--{\bf L1355, TMC2, TMC1, L1552, L1622A-2, L158, 
and  L234E-C}, and possibly {\bf L981-1} and {\bf L1197}.

We propose that these cores in groups 1 and 2 are bona fide infall candidates.

(3) cores with ``significant red excess'' having 
overabundance of red asymmetric CS spectra--
L1521F, L429-1, and CB246.  

(4) cores with  ``no significant excess'' having 
small spread in $\delta V_{CS}$-- L1333, L1495B, L1400A, 
L1517B, L1622A--1, L234E-N, and L1696A. 

(5) cores with ``no significant excess'' having
large spread in $\delta V_{CS}$--
L1495A--N, L1507A, and L1512.

Using our spectroscopic data we determine the main properties of
our nineteen infall candidates: infall radii, speeds, and mass rates. 
We estimate
infall radii from the $\delta  V_{CS}$
distribution and from the distribution of blue-skewed CS spectra, 
finding values typically between 0.06 and 0.14 pc.

The infall speed for each core is derived 
by fitting the spectra with a simple two layer
radiative transfer model, and typical values are
between 0.05 and 0.09~km~s$^{-1}$. 
Finally, approximate mass infall rates are found to range from 
a few $\rm 10^{-6}$ to  $\rm 10^{-5}~\msun~yr^{-1}$.  As these 
infall speeds were obtained using line-of-sight velocities, 
real (3D) values of the infall speed and the mass infall rate should 
be larger than our estimates by a factor of 1--2.
 
A comparison of the equivalent radius of the $\rm N_2H^+$ core 
with the infall radius shows that larger cores tend to have
larger infall radii by a factor of about $2 - 3$, 
suggesting that the infall motions are likely involved
in the process of core condensation.

The parameters of the infall kinematics found 
in the starless cores of our sample 
are very similar to those seen in L1544.
The infall speed and the extent of the infall zone found in our study
can be explained with models of ambipolar diffusion
in super-critical or slightly sub-critical cores, or 
with models of turbulence dissipation,
although further work is needed for a more detailed comparison.
Our observations suggest that extended 
inward motions with large speed are fairly common in starless cores, 
and that such inward motions may be a necessary step
in the condensation of a star-forming dense core.

\acknowledgments
We thank  Tyler Bourke and Jonathan Williams for help during
the observations, and Carl Gottlieb for communicating his
new frequency measurements prior to publication.
We also thank an anonymous referee for helping us to improve
this paper.
C.W.L. acknowledges the financial support by 1996 Overseas 
Postdoctoral Support Program  of 
Korea Research Foundation and partial support by 
``Spectroscopic Observational Study of Celestial 
Object with Optical and Radio Telescopes'' 99-NQ-01-01-A of the 
Korea Astronomy Observatory. C.W.L. also thanks 
the Harvard-Smithsonian Center for Astrophysics
for support while working on this project. 
M.T. acknowledges partial support from Spanish DGES grant PB96-104.
This research was supported  by NASA Origins of 
Solar System Program, Grant NAGW-3401.

\appendix
\label{boloappendix}
\section{Selection and Effects of the New Line Frequencies}
As LMT99 pointed out, in a quantitative study of the infall profiles 
of starless cores with narrow lines, the uncertainty in the 
laboratory determinations of the line frequencies can be a critical issue.
Here we describe the frequency set we have adopted for this
study, and how this choice affects our previous results in LMT99.

In LMT99, we chose a frequency set based on the comparison
of different line profiles in the starless core L1544 (Tafalla et 
al. 1998), and in this way, assumed frequencies of
$97980.950$ MHz for CS (2--1) and $93176.265$ MHz for 
the ``isolated'' component ($\rm F_1F=01-12$) of $\rm N_2H^+~(1-0)$.
This values were at the time more accurate than the existing
laboratory determinations, and were probably limited by the non Gaussian
shape of the line profiles in L1544. Since 
then, a set of more accurate laboratory measurements 
(Gottlieb 2000, private communication) has yielded 
the following frequencies: $97980.953 (\pm 0.002)$ MHz for CS (2--1) and
$109782.173 (\pm 0.002)$ MHz for $\rm C^{18}O~(1-0)$, although 
unfortunately no $\rm N_2H^+$ measurement with equivalent accuracy
has been obtained yet.

The new laboratory measurement of CS (2--1) agrees within the 
errors with the choice in LMT99, so its effect on the LMT99 
CS velocities is minor. Given the 
critical need for our work of an equivalently accurate $\rm N_2H^+$
determination, we have carried out new astronomical observations
of L1512 (the core with narrowest lines) using the IRAM 30m
telescope (Tafalla et al. 2001, in preparation). We have observed 
simultaneously $\rm C^{18}O~(1-0)$ and $\rm N_2H^+$, and used
the new laboratory $\rm C^{18}O~(1-0)$ frequency to determine a value of 
$93176.258$ MHz for the $\rm F_1F=01-12$ component 
of $\rm N_2H^+~(1-0)$. A comparison of the two spectra (assuming the
new frequencies) is shown in Fig. 11.

The new CS(2--1) and $\rm N_2H^+~(1-0)$ frequencies are slightly different 
from those in LMT99 (3 and 7 kHz, respectively), and their use
decreases the velocity shift between the two profiles
by about 0.013 $\rm km~s^{-1}$. The effect of this change on the
$\rm \delta V_{CS}$ distribution is to make it less skewed to the blue.
In Fig. 12 we show how the frequency choice
affects the $\rm \delta V_{CS}$ distribution in the sample 
of starless cores studied by LMT99. 
Cases (a) and (b) are the same as shown in Fig. 6 of LMT99, 
except for a different binning in case (a).
In other words, case (a) uses Lovas (1992)  frequency (97980.968 MHz)
for CS (2--1) and 93176.265 MHz for $\rm N_2H^+$ ``isolated'' component
by Caselli et al. (1995), and case (b)
uses 97980.950 MHz for CS and 93176.265 MHz for $\rm N_2H^+$, as 
preferred by LMT99.
Case (c) shows the $\rm \delta V_{CS}$ distribution with the 
present `best' frequency set
(97980.953 MHz for CS and 93176.258 MHz for $\rm N_2H^+$ ).  
As it can be seen, the new and more accurate set of frequencies
still makes the $\rm \delta V_{CS}$ 
histogram significantly skewed to the blue, so 
the main conclusion in LMT99, that inward motions are a 
significant feature of starless cores, is still valid.
In addition, the new frequency set does not change the list of 
7 strong and 10 probable infall candidates suggested by 
LMT99.

\vfill\eject

\clearpage

\begin{figure}
\begin{center}
{\bf FIGURE CAPTIONS}
\end{center}
\end{figure}

\begin{figure}
\noindent{\bf Fig. 1. ---} Selected CS and $\rm N_2H^+$ emission 
maps of starless cores.
Most figures consist of a CS profile map 
superposed on a $\rm N_2H^+$ integrated intensity map
on the left panel, and of average CS and $\rm N_2H^+$ spectra  
over the half maximum contour of the $\rm N_2H^+$ intensity map 
on the right panel.
The dashed lines on the profiles indicate the
$\rm N_2H^+$ peak velocities. 
For some large sources with multiple cores such as L1495 (Fig. 1-4) 
and L1155C (Fig. 1-20), we present maps of the overall distribution 
of $\rm N_2H^+$ integrated intensity.
L1155C-2 (Fig. 1-21) has a large velocity gradient across the core
($\rm \sim 0.19~km~s^{-1}~arc min^{-1}$), so simply averaging the spectra
will hide any possible infall feature. Thus, for this core we
have shifted the systemic velocity of the spectra 
to make them coincide with that at the brightest $\rm N_2H^+$  position
$\rm (\Delta \alpha, \Delta \delta)=(-88'',-132'')$, and then we
have averaged all spectra within the HM contour of the $\rm N_2H^+$ emission.
The solid lines crossing cores L1495A--N (Fig. 1-5) and 
L1512 (Fig. 1-9) indicate the possible
rotation axis of these cores
(see $\S 3.9$ for further details). Lowest, 
half maximum, and peak levels of the $\rm N_2H^+$ integrated intensity 
($\int T_A^* dv$) for each core are as follows (in $\rm K~km~s^{-1}$); 
(1) L1333 -- 0.108, 0.540, 1.080, (2) L1355 -- 0.051, 0.256, 0.512, 
(3) L1498 -- 0.100, 0.500, 1.000, 
(4) L1495, (5) L1495A--N -- 0.150, 0.750, 1.500;
L1495A--S -- 0.078, 0.390, 0.078, 
(6) L1521F -- 0.13, 0.883, 1.766, 
(7) TMC2 -- 0.163, 0.816, 1.631, (8) TMC1 -- 0.249, 1.245, 2.490,
(9) L1512 -- 0.093, 0.463, 0.926
(10) L1544 -- 0.2, 1.452, 2.904, (11) L1552 -- 0.100, 0.689, 1.377,
(12) L1622A-1 -- 0.096, 0.479, 0.957; L1622A-2 -- 0.100, 0.927, 1.853,
(13) L183 -- 0.160, 0.799, 1.598,
(14) L1689B -- 0.064, 0.322, 0.643,
(15) L158 -- 0.055, 0.274, 0.548, 
(16) L234E-N -- 0.088, 0.442, 0.883; L234E-C -- 0.067, 0.332, 0.672;
L234E-S --  0.094, 0.470, 0.940, (17) L492 -- 0.142, 0.711, 1.421,
(18) L429-1 -- 0.178, 0.891, 1.782,
(19) L694-2 -- 0.167, 0.835, 1.670, (20) L1155C,
(21) L1155C-2 -- 0.070, 0.348, 0.695, (22) L1155C-1 -- 0.075, 0.374, 0.748,
(23) L981-1 -- 0.045, 0.225, 0.449, (24) L1197 -- 0.073, 0.365, 0.729,
 
\end{figure}

\begin{figure}
\noindent{\bf Fig. 2. --- } 
$\rm N_2H^+$ line width versus size for the $\rm N_2H^+$ cores in our sample.
The core size is the equivalent radius of the half maximum 
(HM) contour
of the $\rm N_2H^+$ integrated map, and the line width 
is the average of the FWHMs of the 
line profiles whose peak S/N is better than 5.
Filled squares, triangles, and open circles represent 
strong infall candidates, possible infall candidates, and 
all of non-infall candidates, respectively, which we
classify in \S 3-4.
In the figure we exclude six cores
(CB23, L1689B, L429--1, L1148B, L981--1, and L1063) because of 
the large uncertainty in their $\rm <\Delta V_{N_2H^+}>$
($\rm \sigma_{<\Delta V_{N_2H^+}>} \ga 0.03~km~s^{-1}$).

\end{figure}

\begin{figure}
\noindent{\bf Fig. 3. --- } Number distributions of $<\delta V_{CS}>$
for our sample of starless cores. This histogram shows that most of our 
sources have a global
overabundance of blue spectra characteristic of inward motions.
Note, however, that this does not necessarily reflect 
a statistical excess of infall motions in starless cores,
because many sources  in our present sample have been selected with 
a bias toward infall candidates found in previous 
observations.
For L183, we use a $<\delta V_{CS}>$ from the spectra within 
the HM contour of the $\rm N_2H^+$ emission. 

\end{figure}

\begin{figure}
\noindent{\bf Fig. 4. --- } Mean of 
$\delta V_{CS}$ ($<\delta V_{CS}>$) versus blue excess (E) 
for each source. Note the good correlation.
The error bar of $\delta V_{CS}$ is 1 standard error of the mean
$<\delta V_{CS}>$.
For L183, we use spectra within the HM contour of the $\rm N_2H^+$ 
map (see text).

\end{figure}

\begin{figure}
\noindent{\bf Fig. 5. --- } Mean of $\delta V_{CS}$ ($<\delta V_{CS}>$)
versus probability of the t-test (P). Note how P approaches zero as 
$<\delta V_{CS}>$ deviates from zero.
The error bar is 1 standard error of the mean
$<\delta V_{CS}>$. For L183, we use only spectra within  
the HM contour of the $\rm N_2H^+$ map (see text).
\end{figure}

\begin{figure}
\noindent{\bf Fig. 6. --- }  Number distributions of $ \delta V_{CS}$
for cores in our five groups; 
(a) group 1 -- cores with  ``very significant blue excess'' having 
overwhelming overabundance of blue asymmetric CS spectra
with {\it double} peaks -- ``strong infall candidates'',
(b) group 2 -- cores having similar spectral properties 
as those in the group 1, and having {\it single}-peaked CS spectra
which are skewed to the blue blue -- ``probable infall candidates'',
(c) group 3 -- cores with ``significant red excess'' having 
overwhelming overabundance of red asymmetric CS spectra,
(d) group 4 -- cores with  ``no significant excess'' 
having a small spread in $\delta V_{CS}$, and
(e) cores with ``no significant excess'' having a large spread 
in $\delta V_{CS}$. All histograms are normalized by the  
total number of samples. For L183, we use only spectra within
the HM contour of the $\rm N_2H^+$ map (see text).

\end{figure}

\begin{figure}
\noindent{\bf Fig. 7. --- } Number distributions of infall radii 
for (a) weak, (b) strong cases determined from the $\delta V_{CS}$ 
distribution, and (c) case determined from the distribution of 
blue-skewed CS profiles. 
See text in $\S 3.5$ for a more detailed explanation of the three cases.

\end{figure}

\begin{figure}
\noindent{\bf Fig. 8. ---} Best fit results of a two-layer radiative 
transfer model to  observed spectra.
Model spectra (thick solid line) are superposed on
the observed spectra (thin histogram). 
W and S mean the `weak'
and `strong' cases, respectively, as described in  $\S 3.6$.
Each observed spectrum is an average of spectra satisfying 
each case, except for the S case in  L1155C--2, 
where the spectrum at $(-88'',-176'')$ is presented. 
For the S case in L234E--S, one spectrum (out of 3) was discarded 
because its systemic velocity is slightly 
different from that of the other spectra.
 
\end{figure}

\begin{figure}
\noindent{\bf Fig. 9. ---} Number distribution of infall speeds
for (a) `weak' and (b) `strong' cases 
\end{figure}

\begin{figure}
\noindent{\bf Fig. 10. ---} Comparison of 
the infall radius with the size of the $\rm N_2H^+$ emission.
Three cases are presented: (a) weak case,  
(b) strong case from the $\delta V_{CS}$ distribution, and (c) 
case determined from the distribution of blue-skewed CS profiles.
L183 has not been included because this core is assumed to
be a strong infall candidate from a statistics of the spectra within 
the HM contour of the $\rm N_2H^+$ map, so its infall radius 
was simply assumed to be equal to the equivalent radius of the HM contour.
\end{figure}

\begin{figure}
\noindent{\bf Fig. 11. ---} Comparison of $\rm C^{18}O~(1-0)$ 
and $\rm N_2H^+~(1-0)$ $\rm F_1F=01-12$ toward L1512, the core with
the narrowest lines. For $\rm C^{18}O~(1-0)$, we have used the 
the recent laboratory determination by C.Gottlieb (priv. comm.): 
$109782.173(\pm 0.002)$ MHz. For $\rm N_2H^+~(1-0)$ $\rm F_1F=01-12$,
we have derived a frequency of $93176.258$ MHz from the best match
with $\rm C^{18}O~(1-0)$.
\end{figure}

\begin{figure}
\noindent{\bf Fig. 12. ---}
Effect of the choice of frequencies for CS (2--1) and  $\rm N_2H^+~(1-0)$
on the distribution of $\rm \delta V_{CS}$ in starless cores (LMT99 data).
Fig. 1-(a) and (b) are the same as Fig. 6 of LMT99, and Fig. 1-(c) 
shows the result of using our new set of frequencies. Note how the new 
frequencies make the $\rm \delta V_{CS}$ histogram still 
significantly skewed to the blue.
\end{figure}

\end{document}


\begin{deluxetable}{lrrccccc}
\tablecaption{ Starless core regions mapped in 
$\rm CS~(2-1)$, $\rm N_2H^+~(1-0)$, and $\rm C^{18}O~(1-0)$.\tablenotemark{a}}
\tablehead{
        \colhead{Source}&    
        \colhead{RA 1950} & \colhead{DEC 1950}     &
        \colhead{Array} & \colhead{CS} &
        \colhead{$\rm N_2H^+$}&\colhead{$\rm C^{18}O$} & 
        \colhead{selection} 
        \nl
        \colhead{} & \colhead{$^h~~^m~~^s~$}&\colhead{~~~~$^{\circ}~~~'~~~''$} &
        \colhead{} & \colhead{} & \colhead{} & \colhead{} &
        \colhead{basis}
        }
\startdata
  L1333     &02~21~10.1 & 75~13~31& S    & D     & D     &\nodata & 3    \nl
  L1355     &02~48~41.4 & 68~43~36& S(Q) & D     & D     &\nodata & 1    \nl
  L1498     &04~07~50.0 & 25~02~13& S(Q) & D     & D     & D      & 1    \nl
  L1495     &04~15~25.5 & 28~19~31& S(Q) & D     & D     &\nodata & 5    \nl
  L1521B    &04~21~08.5 & 26~30~00& Q    & D     &ND     & D      & 4    \nl
  L1521F    &04~25~35.0 & 26~45~00& S(Q) & D     & D     &\nodata & 1    \nl
  L1400FD   &04~25~52.6 & 54~05~47& Q    & D     &ND     & D      & 5    \nl
  L1521E    &04~26~17.0 & 26~07~47& Q    & D     &ND     & D      & 4    \nl
  L1400A    &04~26~50.6 & 54~46~08& S(Q) & D     & D     & D      & 3    \nl
  L1445     &04~28~28.8 & 46~31~01& S    & D     &ND     & D      & 2    \nl
  TMC2      &04~29~46.9 & 24~18~54& S(Q) & D     & D     & D      & 2    \nl
  ON32      &04~36~24.0 & 25~45~00& Q    & D     & D     &\nodata & 5    \nl
  TMC1      &04~38~29.0 & 25~39~01& S(Q) & D     & D     &\nodata & 4    \nl
  ON34      &04~38~42.0 & 25~36~00& S(Q) & D     & D     &\nodata & 5    \nl
  L1507A    &04~39~29.0 & 29~38~07& Q    & D     & D     &\nodata & 3    \nl
  CB23      &04~40~22.0 & 29~34~26& Q    & D     & D     &\nodata & 2    \nl
  L1517C    &04~51~35.9 & 30~30~00& Q    & D     &ND     & D      & 4    \nl
  L1517A    &04~51~54.8 & 30~28~53& Q    & D     & D     & D      & 3    \nl
  L1517B    &04~52~07.2 & 30~33~18& Q    & D     & D     & D      & 3    \nl
  L1512     &05~00~54.4 & 32~39~00& S(Q) & D     & D     & D      & 3    \nl
  L1544     &05~01~14.0 & 25~07~00& S(Q) & D     & D     &\nodata & 2    \nl
  L1552     &05~14~34.9 & 26~02~07& S(Q) & D     & D     &\nodata & 3    \nl
  L1622     &05~52~02.9 & 01~53~15& S(Q) & D     & D     &\nodata & 2    \nl
  L134A     &15~50~58.1 &-04~26~36& Q    & D     & D     & D      & 3    \nl
  L183      &15~51~30.0 &-02~42~51& S(Q) & D     & D     & D      & 1    \nl
  L1696A    &16~25~29.9 &-24~12~32& Q    & D     & D     & D      & 3    \nl
  L1696B    &16~26~01.5 &-24~14~59& Q    & D     & D     & D      & 2    \nl
  L1709A    &16~27~50.0 &-23~34~37& Q    & D     & ND    & D      & 3    \nl
  L1709C    &16~30~52.5 &-23~32~18& Q    & D     &\nodata& D      & 4    \nl
  L1689B    &16~31~43.6 &-24~31~40& S(Q) & D     & D     & D      & 2    \nl
  L158      &16~44~37.7 &-13~54~03& S    & D     & D     &\nodata & 1    \nl
  L191      &16~44~41.5 &-12~23~20& Q    & D     & ND    & D      & 4    \nl
  L204C     &16~44~51.0 &-12~02~47& Q    & D     &\nodata& D      & 4    \nl
  L204B     &16~45~00.0 &-11~50~00& Q    & D     &\nodata& D      & 4    \nl
  L204F     &16~45~01.3 &-11~51~40& Q    & D     & ND    & D      & 4    \nl
  L234E     &16~45~22.6 &-10~52~10& S    & D     & D     &\nodata & 2    \nl
  L63       &16~47~21.0 &-18~01~00& Q    & D     &\nodata& D      & 3    \nl
  L492      &18~13~11.6 &-03~47~17& S    & D     & D     &\nodata & 2    \nl
  L429-1    &18~14~28.2 &-08~14~40& S    & D     & D     &\nodata & 3    \nl
  L694-2    &19~38~42.4 & 10~49~59& S    & D     & D     &\nodata & 1    \nl
  L1155C    &20~43~00.0 & 67~41~47& S(Q) & D     & D     & D      & 1    \nl
  L1147     &20~40~00.0 & 67~11~00& Q    & D     & ND    & D      & 4    \nl
  L1148     &20~40~38.8 & 67~09~48& S    &\nodata& D     &\nodata & 3    \nl
  L1082C    &20~50~19.6 & 60~07~15& Q    & D     &\nodata&\nodata & 4    \nl
  L981-1    &20~58~36.3 & 50~08~19& S    & D     & D     &\nodata & 3    \nl
  L1063     &21~06~44.6 & 56~06~09& S    & D     & D     &\nodata & 3    \nl
  L1235B    &22~14~41.2 & 73~07~31& Q    & D     &\nodata& D      & 5    \nl
  L1235     &22~14~41.4 & 73~09~25& Q    & D     & ND    &\nodata & 4    \nl
  L1197     &22~35~08.1 & 58~41~45& S    & D     & D     &\nodata & 3    \nl
  L1103     &22~40~14.3 & 56~27~58& Q    &ND     & ND    &ND      & 5    \nl
  L1234     &23~16~08.4 & 62~12~08& Q    &ND     &\nodata&\nodata & 3    \nl
  CB244     &23~25~00.5 & 74~01~06& Q    & D     &\nodata& D      & 5    \nl
  CB246     &23~54~12.0 & 58~17~47& Q(S) & D     & D     & D      & 3    \nl
\enddata
\tablenotetext{a}{
The first column indicates the name of
the observed region, and the second and third columns 
list the central coordinates in B1950 epoch. 
The 4th column designates the receiver array, QUARRY (Q) and SEQUOIA (S),
and `S(Q)' means that both SEQUOIA and
QUARRY observations were carried out. `D', `ND', and `\nodata' in the
5th to 7th columns indicate `Detection', `No Detection', and `Not Observed',
respectively. The last column indicates the selection
criterion for each source:
1 \& 2 -- selected as strong or probable infall candidate according to LMT99, 
3 -- selected from strong CS \& $\rm N_2H^+$ detection according to LMT99, 
4 -- selected from $\rm NH_3$ detection according to BM89, and 
5 -- selected from LM99 catalog.}
\end{deluxetable}

\special{landscape}
\begin{deluxetable}{lrrlrrrrlrc}
\small
\tablecaption{ Starless $\rm N_2H^+$ Cores, and 
Their Velocity and Map Information.\tablenotemark{a}}
\tablehead{
        \colhead{Source}&    
        \colhead{RA 1950}       & \colhead{DEC 1950}      & \colhead{$({A \over \pi})^{1/2}$} &
        \colhead{$<V_{CS}>$}     & \colhead{$<V_{N_2H^+}>$}     &
        \colhead{$<\Delta V_{N_2H^+}>$} & \colhead{$\rm N$} & 
        \colhead{ R} & \colhead{$<\delta V_{CS}>$} & \colhead{$<\tau_{N_2H^+}>$} \nl
        \colhead{}      &
        \colhead{$^h~~^m~~^s~$} & \colhead{~~~~$^{\circ}~~~'~~~''$}& \colhead{(pc)} &
        \colhead{($\rm km~s^{-1}$)}   & \colhead{($\rm km~s^{-1}$)} & \colhead{($\rm km~s^{-1}$)} &
        \colhead{} & \colhead{(pc)} & \colhead{}  & \colhead{}
        }
\startdata
L1333    &02~21~10.1 & 75~13~31&($0.76'$)&  3.84$\pm$0.06&  3.86$\pm$0.05&0.34$\pm$0.02& 5&($1.20'$)&-0.07$\pm$0.05&1.0$\pm$0.1\nl
L1355    &02~48~41.4 & 68~43~36&  0.05   & -3.89$\pm$0.02& -3.77$\pm$0.02&0.25$\pm$0.03& 4&  0.05   &-0.48$\pm$0.07&0.2$\pm$0.1\nl
L1498    &04~07~50.0 & 25~02~13&  0.05   &  7.69$\pm$0.01&  7.81$\pm$0.01&0.20$\pm$0.01&15&  0.08   &-0.66$\pm$0.04&0.7$\pm$0.1\nl   
L1495B   &04~14~59.0 & 28~15~05&  0.04   &  6.57$\pm$0.02&  6.55$\pm$0.04&0.37$\pm$0.03& 3&  0.03   & 0.04$\pm$0.13&0.5$\pm$0.2\nl
L1495A-N &04~15~25.5 & 28~20~15&  0.05   &  7.36$\pm$0.05&  7.30$\pm$0.01&0.33$\pm$0.01&18&  0.11   & 0.25$\pm$0.15&0.5$\pm$0.1\nl
L1495A-S &04~15~35.6 & 28~16~35&  0.06   &  7.19$\pm$0.02&  7.31$\pm$0.01&0.26$\pm$0.01&14&  0.14   &-0.45$\pm$0.09&0.8$\pm$0.2\nl   
L1521F   &04~25~35.0 & 26~45~00&  0.04   &  6.62$\pm$0.01&  6.48$\pm$0.01&0.27$\pm$0.01&10&  0.07   & 0.53$\pm$0.04&1.0$\pm$0.1\nl
L1400A   &04~26~55.7 & 54~46~08&  0.06   &  3.22$\pm$0.03&  3.20$\pm$0.03&0.25$\pm$0.01&12&  0.10   & 0.10$\pm$0.05&0.6$\pm$0.2\nl
TMC2     &04~29~46.9 & 24~18~54&  0.06   &  6.05$\pm$0.01&  6.13$\pm$0.02&0.33$\pm$0.01&40&  0.13   &-0.24$\pm$0.03&0.6$\pm$0.1\nl
TMC1     &04~38~29.0 & 25~39~01&  0.05   &  5.72$\pm$0.02&  5.89$\pm$0.01&0.32$\pm$0.01&28&  0.15   &-0.52$\pm$0.05&0.5$\pm$0.1\nl
L1507A   &04~39~29.0 & 29~38~07&  0.04   &  6.14$\pm$0.04&  6.18$\pm$0.01&0.22$\pm$0.01& 3&  0.05   &-0.17$\pm$0.20&0.2$\pm$0.1\nl
CB23     &04~40~18.2 & 29~33~36&  0.03   &  6.04$\pm$0.02&  6.02$\pm$0.02&0.22$\pm$0.03& 0&\nodata  & 0.14$\pm$0.17&\nodata    \nl
L1517B   &04~52~07.2 & 30~33~18&  0.04   &  5.82$\pm$0.02&  5.81$\pm$0.01&0.21$\pm$0.02& 4&  0.04   & 0.04$\pm$0.07&0.9$\pm$0.2\nl
L1512    &05~00~50.9 & 32~39~00&  0.04   &  7.02$\pm$0.04&  7.06$\pm$0.02&0.19$\pm$0.01&10&  0.07   &-0.22$\pm$0.16&0.4$\pm$0.1\nl
L1544    &05~01~10.8 & 25~07~00&  0.03   &  7.08$\pm$0.08&  7.14$\pm$0.02&0.28$\pm$0.02& 5&  0.05   &-0.24$\pm$0.26&0.8$\pm$0.2\nl
L1552    &05~14~31.6 & 25~02~07&  0.03   &  7.60$\pm$0.01&  7.69$\pm$0.01&0.21$\pm$0.00& 7&  0.06   &-0.46$\pm$0.03&1.0$\pm$0.2\nl
L1622A-2 &05~52~02.9 & 01~53~15&  0.11   &  1.03$\pm$0.02&  1.09$\pm$0.01&0.28$\pm$0.01&10&  0.27   &-0.22$\pm$0.06&0.7$\pm$0.2\nl
L1622A-1 &05~52~17.5 & 01~56~56&  0.15   &  1.18$\pm$0.02&  1.15$\pm$0.03&0.25$\pm$0.01&10&  0.25   & 0.13$\pm$0.07&0.5$\pm$0.1\nl
L134A    &15~50~54.8 &-04~26~36&  0.05   &  2.76$\pm$0.01&  2.72$\pm$0.04&0.23$\pm$0.01& 2&  0.08   & 0.16$\pm$0.21&\nodata    \nl
L183     &15~51~29.9 &-02~43~35&  0.11   &  2.34$\pm$0.02&  2.37$\pm$0.01&0.26$\pm$0.01&55&  0.28   &-0.06$\pm$0.07&0.7$\pm$0.1\nl   
         &           &         &         &  2.32$\pm$0.02&  2.40$\pm$0.01&0.25$\pm$0.01&20&         &-0.32$\pm$0.09&0.9$\pm$0.1\nl   
L1696A   &16~25~29.9 &-24~12~32&  0.04   &  3.42$\pm$0.02&  3.39$\pm$0.03&0.21$\pm$0.01& 3&  0.04   & 0.15$\pm$0.16&0.3$\pm$0.2\nl
L1689B   &16~31~43.6 &-24~31~40&  0.04   &  3.44$\pm$0.04&  3.54$\pm$0.02&0.32$\pm$0.03& 8&  0.09   &-0.30$\pm$0.11&0.2$\pm$0.1\nl   
L158     &16~44~33.7 &-13~53~19&  0.04   &  3.79$\pm$0.03&  3.91$\pm$0.01&0.21$\pm$0.01& 6&  0.07   &-0.60$\pm$0.09&0.3$\pm$0.1\nl
L234E-N  &16~45~16.2 &-10~41~53&  0.05   &  2.88$\pm$0.02&  2.92$\pm$0.02&0.19$\pm$0.01& 8&  0.08   &-0.16$\pm$0.07&0.8$\pm$0.2\nl
L234E-C  &16~45~16.1 &-10~44~05&  0.03   &  2.88$\pm$0.06&  2.92$\pm$0.01&0.22$\pm$0.01& 3&  0.04   &-0.18$\pm$0.20&0.1$\pm$0.1\nl
L234E-S  &16~45~22.6 &-10~52~10&  0.04   &  2.89$\pm$0.03&  3.05$\pm$0.02&0.26$\pm$0.01& 6&  0.06   &-0.63$\pm$0.14&0.6$\pm$0.2\nl   
L492     &18~13~07.9 &-03~47~16&  0.04   &  7.63$\pm$0.05&  7.74$\pm$0.01&0.31$\pm$0.02& 7&  0.09   &-0.35$\pm$0.21&1.0$\pm$0.3\nl   
L429-1   &18~14~22.2 &-08~14~39&  0.05   &  7.01$\pm$0.07&  6.82$\pm$0.02&0.42$\pm$0.03&12&  0.11   & 0.47$\pm$0.16&0.4$\pm$0.1\nl
L694-2   &19~38~42.4 & 10~49~59&  0.04   &  9.41$\pm$0.01&  9.59$\pm$0.02&0.25$\pm$0.01& 7&  0.09   &-0.73$\pm$0.08&0.7$\pm$0.2\nl   
L1155C-2 &20~42~28.6 & 67~37~25&  0.15   &  1.30$\pm$0.06&  1.44$\pm$0.06&0.30$\pm$0.02&12&  0.31   &-0.51$\pm$0.05&0.5$\pm$0.1\nl   
L1155C-1 &20~43~00.0 & 67~41~47&  0.13   &  2.54$\pm$0.01&  2.70$\pm$0.02&0.38$\pm$0.02& 6&  0.15   &-0.45$\pm$0.04&0.4$\pm$0.2\nl   
L1148A*  &20~40~38.8 & 67~09~48&  0.06   &               &  2.59$\pm$0.01&0.24$\pm$0.02& 4&  0.08   &              &0.2$\pm$0.1\nl
L1148B*  &20~40~31.4 & 67~12~00&  0.04   &               &  2.58$\pm$0.03&0.28$\pm$0.06& 1&  0.04   &              & \nodata   \nl
L981-1   &20~58~36.4 & 50~09~03&  0.12   & -0.20$\pm$0.01& -0.06$\pm$0.01&0.36$\pm$0.04& 1&  0.09   &-0.41$\pm$0.07&0.6$\pm$0.3\nl
L1063    &21~06~44.6 & 56~06~09&  0.13   & -0.94$\pm$0.02& -0.86$\pm$0.02&0.36$\pm$0.05& 1&  0.10   &-0.23$\pm$0.09&1.2$\pm$0.6\nl
L1197    &22~35~08.1 & 58~41~45&  0.06   & -3.20$\pm$0.03& -3.12$\pm$0.02&0.23$\pm$0.00& 3&  0.10   &-0.37$\pm$0.04&0.5$\pm$0.1\nl
CB246    &23~54~17.5 & 58~17~47&  0.04   & -0.76$\pm$0.03& -0.83$\pm$0.01&0.28$\pm$0.01& 7&  0.05   & 0.28$\pm$0.09&0.5$\pm$0.1\nl
\enddata                                                                                                       
\tablenotetext{a}{This table lists the thirty 
seven $\rm N_2H^+$ cores for which the spatial 
extent and the strongest positions are well defined.
L1148A and B have no $\rm \delta V_{CS}$ information because 
they were observed in $\rm N_2H^+~(1-0)$ only.
The first column indicates the name of the $\rm N_2H^+$ core, 
and the second and the third columns 
list the coordinates in B1950 epoch of the 
positions of $\rm N_2H^+$  peak intensity, 
which are sometimes slightly different from those of the map centers given 
in Table 1. Column 4 lists an equivalent radius $\rm ({A / \pi})^{1/2}$ 
of the $\rm N_2H^+$ emission, where A is an area inside the half 
maximum contour of $\rm N_2H^+$ integrated emission. The velocities and their 
$\rm <\delta V_{CS}>$ given in columns 5 to 7 and column 10 are averages of 
Gaussian fit velocities and $\rm \delta V_{CS}$s of the 
spectra whose peak S/N ratio is better than 5:
$\rm <V_{CS}>={\sum_i V_{CS}^i /N}$, 
$\rm <V_{N_2H^+}>={\sum_i V_{N_2H^+}^i /N}$, 
$\rm <\Delta V_{N_2H^+}>={\sum_i \Delta V_{N_2H^+}^i /N}$, and
$\rm <\delta V_{CS}>={\sum_i \delta V_{CS}^i /N}$, where
$\delta V_{CS} = (V_{CS}-V_{N_2H^+})/ \Delta V_{N_2H^+}$. 
CB23 has no $\rm \delta V_{CS}$ spectra detected with S/N better than 5, 
so the values for this source 
are averages from spectra with marginal detection. 
All estimates are given with the standard error of the mean (s.e.m).  
N in column 8 is the number of samples for which the peak S/N of 
the spectra is larger than 5. 
`R' in column 9 is one half the largest separation between positions 
having spectra with peak S/N larger than 5. The 
core distances used to derive linear radii are mainly 
from Lee \& Myers (1999) and references therein.
We assume that the distance to L694-2 is the same as to 
B335 (250 pc) because L694-2 lies within a few degrees of B335 and has a
similar LSR velocity. The distance to the L1197 was taken as 
400 pc following Jijina, Myers, \& Adams (1999). The spatial
extent for L1333 is given in angular scale due to its unknown 
distance.  $<\tau_{N_2H^+}>$ in column 11 is the optical depth of the 
isolated ($\rm F_1F=01-12$) hyperfine component in a core map, defined 
as one-ninth of the total optical depth of the $1-0$ spectrum, averaged 
over all spectra in a map having S/N $>5$. For L183, we give two 
measurements: the first row uses the spectra with
peak S/N larger than 5 and the second row uses only
spectra within the HM contour of the $\rm N_2H^+$ map.} 
\end{deluxetable}                                                                                              

\special{landscape}
\begin{deluxetable}{lrrrrrr}
\tablecaption{$\delta V_{CS}$ related parameters of Starless $\rm N_2H^+$ Cores.\tablenotemark{a}}
\tablehead{
        \colhead{Source}&    
        \colhead{$\rm N_-$}     & \colhead{$\rm N_+$}     &
        \colhead{$\rm N_{0}$} & \colhead{$\rm N$} & 
        \colhead{ E} & 
        \colhead{P} \nl
        \colhead{}      &
        \colhead{}     & \colhead{}     & \colhead{} &
        \colhead{} & \colhead{} &
        \colhead{}
        }
\startdata
L1333    &   0    &  0     &  11   &   12      &    0.00     &   0.30        \nl
L1355    &   5    &  0     &   7   &   12      &    0.42     &   0.00        \nl
L1498    &  14    &  0     &   6   &   20      &    0.70     &   0.00        \nl
L1495B   &   0    &  1     &   6   &    7      &   -0.14     &   0.80        \nl
L1495A-N &  11    & 11     &   7   &   29      &    0.00     &   0.17        \nl
L1495A-S &  11    &  0     &  15   &   26      &    0.42     &   0.00        \nl
L1521F   &   0    & 10     &   4   &   14      &   -0.71     &   0.00        \nl
L1400A   &   0    &  0     &  24   &   24      &    0.00     &   0.02        \nl
TMC2     &  25    &  0     &  24   &   49      &    0.51     &   0.00        \nl
TMC1     &  26    &  0     &  16   &   42      &    0.62     &   0.00        \nl
L1507A   &   0    &  0     &   9   &    9      &    0.00     &   0.08        \nl
CB23     &   0    &  0     &   4   &    4      &    0.00     &   0.47        \nl
L1517B   &   0    &  0     &   7   &    7      &    0.00     &   0.93        \nl
L1512    &   4    &  3     &   7   &   14      &    0.07     &   0.12        \nl
L1544    &   7    &  1     &   9   &   17      &    0.35     &   0.00        \nl
L1552    &   7    &  0     &   6   &   13      &    0.54     &   0.00        \nl
L1622A-2 &   5    &  2     &  10   &   17      &    0.18     &   0.08        \nl
L1622A-1 &   0    &  2     &  10   &   13      &   -0.15     &   0.82        \nl
L134A    &   0    &  0     &   9   &    9      &    0.00     &   0.00        \nl
L183     &  15(13)&  3(1)  &  52(6)&   76(20)  &    0.12(20) &   0.62(0.00)  \nl
L1696A   &   0    &  0     &   7   &    7      &    0.00     &   0.35        \nl
L1689B   &   7    &  2     &   5   &   14      &    0.36     &   0.01        \nl
L158     &   2    &  0     &  11   &   13      &    0.15     &   0.00        \nl
L234E-N  &   0    &  0     &  11   &   11      &    0.00     &   0.01        \nl
L234E-C  &   1    &  0     &   7   &    8      &    0.13     &   0.01        \nl
L234E-S  &   4    &  0     &   6   &   10      &    0.40     &   0.00        \nl
L492     &   5    &  1     &   2   &    8      &    0.50     &   0.10        \nl
L429-1   &   4    &  9     &   3   &   16      &   -0.31     &   0.02        \nl
L694-2   &   8    &  1     &   3   &   12      &    0.58     &   0.01        \nl
L1155C-2 &   9    &  0     &   9   &   18      &    0.50     &   0.00        \nl
L1155C-1 &   4    &  0     &   5   &    9      &    0.44     &   0.00        \nl
L1148A   &\nodata &\nodata &\nodata&\nodata    &\nodata      &\nodata        \nl
L1148B   &\nodata &\nodata &\nodata&\nodata    &\nodata      &\nodata        \nl
L981-1   &   2    &  0     &   1   &    3      &    0.67     &   0.08        \nl
L1063    &   0    &  0     &   1   &    1      &    0.00     & \nodata       \nl
L1197    &   1    &  0     &   3   &    4      &    0.25     &   0.05        \nl
CB246    &   0    &  2     &   9   &   11      &   -0.18     &   0.03        \nl
\enddata                              									       

\tablenotetext{a}{ $\rm N_-$ is the number of positions with
$\rm \delta V_{CS} \le -5~\sigma_{\delta V_{CS}}$, $\rm N_+$ 
the number of positions with $\rm \delta V_{CS} \ge 5~\sigma_{\delta V_{CS}}$, 
$\rm N_0$ the number of positions with $\rm  -5~\sigma_{\delta V_{CS}}
< \delta V_{CS} <  5~\sigma_{\delta V_{CS}}$, and $\rm N=N_- + N_+ + N_0$. 
E in column 6 is the `blue excess,' defined as $\rm E={(N_- - N_+)/ N}$. 
`P' in the last column is the probability of drawing the observed 
$ \delta V_{CS}$ distribution from a zero mean t-distribution, 
calculated from a student t-test. For L183, we give two measurements using
all spectra in the $\rm N_2H^+$ core
(given without parenthesis) and using only spectra
within the HM contour of the $\rm N_2H^+$ emission
(bracketed with parenthesis).}

\end{deluxetable}											       

\special{landscape}
\begin{deluxetable}{lccccrccccc}
\small
\tablecaption{Results of Line Profile Model for 19 Starless Cores.\tablenotemark{a}}

\tablehead{
        \colhead{Source}& \colhead{Case}&   
        \colhead{$\tau_0$}       & \colhead{$\sigma$ }      &
        \colhead{$\rm T_r$}      &
        \colhead{$\rm V_{thin}$} &\colhead{$\rm V_{in,z}$}& 
        \colhead{$\rm R_{in}(\delta V_{CS})$} & \colhead{$\rm R_{in}(CS)$} & 
        \colhead{n} & \colhead{$\rm \dot M$} \nl
        \colhead{}&     
        \colhead{} &     \colhead{} & \colhead{($\rm km~s^{-1}$)}      &
        \colhead{(K)}     &
        \colhead{($\rm km~s^{-1}$)} &\colhead{($\rm km~s^{-1}$)} & 
        \colhead{(pc)}& \colhead{(pc)} & \colhead{($\times 10^4~cm^{-3}$)} & 
        \colhead{($\times 10^{-5}\msun ~year^{-1}$)}
        }
\startdata
L1355     &    W  &   0.68  &0.14 &   7.9 & -3.75 &  0.121&    0.11& 0.20 &       0.91 &   0.84  \nl
L1498     &    W  &   1.53  &0.09 &   7.0 &  7.79 &  0.044&    0.08& 0.08 &       0.71 &   0.13  \nl
          &    S  &   1.55  &0.09 &   7.8 &  7.78 &  0.030&    0.05&      &       1.82 &   0.09  \nl
L1495A-S  &    W  &   1.33  &0.20 &   5.1 &  7.36 &  0.078&    0.09& 0.12 &       2.78 &   1.11  \nl 
          &    S  &   1.33  &0.20 &   5.1 &  7.36 &  0.072&    0.09&      &       2.78 &   1.03  \nl
TMC2      &    W  &   1.04  &0.26 &   7.8 &  6.40 &  0.165&    0.13& 0.14 &       2.25 &   3.97  \nl
TMC1      &    W  &   0.60  &0.18 &  12.2 &  5.77 &  0.052&    0.17& 0.16 &       0.63 &   0.60  \nl
L1544     &  W(S) &   3.26  &0.11 &   9.2 &  7.19 &  0.021&    0.07& 0.09 &       1.39 &   0.09  \nl
L1552     &    W  &   1.60  &0.14 &   7.1 &  7.78 &  0.116&    0.07& 0.12 &       2.25 &   0.81  \nl
L1622A-2  &    W  &   1.18  &0.19 &   9.0 &  1.21 &  0.093&    0.23& 0.24 &       0.30 &   1.20  \nl
L183      &    S  &   1.27  &0.24 &   6.3 &  2.57 &  0.048&    0.11&      &       2.68 &   0.98  \nl
L1689B    &    W  &   1.56  &0.17 &  10.1 &  3.60 &  0.050&    0.08& 0.18 &       2.54 &   0.51  \nl
          &    S  &   1.57  &0.17 &  10.1 &  3.97 &  0.048&    0.08&      &       2.54 &   0.49  \nl
L158      &    W  &   1.46  &0.17 &   5.0 &  3.98 &  0.074&    0.02& 0.11 &      40.61 &   0.76  \nl
L234E-C   &    W  &   0.97  &0.14 &   9.2 &  2.89 &  0.047&    0.02& 0.07 &      27.54 &   0.33  \nl
L234E-S   &    W  &   0.98  &0.17 &  10.9 &  3.03 &  0.054&    0.06& 0.12 &       4.51 &   0.56  \nl
          &    S  &   1.22  &0.15 &  12.1 &  3.06 &  0.050&    0.06&      &       3.51 &   0.40  \nl
L492      &    W  &   0.78  &0.21 &   8.8 &  7.72 &  0.086&    0.07& 0.09 &       5.06 &   1.35  \nl
          &    S  &   1.38  &0.17 &   7.9 &  7.70 &  0.043&    0.05&      &       6.50 &   0.44  \nl
L694-2    &    W  &   1.73  &0.16 &   5.9 &  9.63 &  0.074&    0.12& 0.13 &       1.00 &   0.67  \nl
          &    S  &   2.08  &0.16 &   6.2 &  9.64 &  0.054&    0.09&      &       1.78 &   0.49  \nl
L1155C-2  &    W  &   1.19  &0.18 &   6.1 &  1.24 &  0.100&    0.22& 0.45 &       0.38 &   1.15  \nl
          &    S  &   1.69  &0.13 &   6.0 &  1.62 &  0.065&    0.12&      &       0.66 &   0.39  \nl
L1155C-1  &    W  &   1.47  &0.27 &   5.7 &  2.87 &  0.096&    0.12& 0.25 &       2.85 &   2.49  \nl
          &    S  &   1.50  &0.27 &   5.6 &  2.86 &  0.089&    0.09&      &       5.06 &   2.31  \nl
L981-1    &  W(S) &   0.90  &0.18 &   5.3 & -0.04 &  0.098&    0.19& 0.31 &       0.50 &   1.13  \nl
L1197     &    W  &   1.45  &0.11 &   8.0 & -3.11 &  0.050&    0.05& 0.10 &       2.72 &   0.22  \nl
\enddata                              				       					       
\tablenotetext{a}{
This table presents infall parameters for 19 infall candidates. 
The first column indicates the name of the $\rm N_2H^+$ core. 
In the second column, `W' means the `weak' case considering all blue
asymmetric spectra satisfying  
$\rm \delta V_{CS} \le -5~\sigma_{\delta V_{CS}}$, `S' means 
the `strong' case considering only those CS spectra 
with double-peaked infall asymmetry satisfying 
$\rm \delta V_{CS} \le -5~\sigma_{\delta V_{CS}}$, and `W(S)' means 
that the W and  S cases produce the same average spectrum.  
If no `S' is listed, there are no double-peaked spectra having S/N $>5$.
The $\tau_0$ (the peak optical depth) in the 3rd column,
the $\sigma$ (the velocity dispersion) in the 4th column,
the $\rm T_r$ (the excitation temperature of the rear layer) in the 5th column,
the $\rm V_{thin}$ (the peak velocity
of the optically thin spectrum) in the 6th column, and
the $\rm V_{in,z}$ (one dimensional infall speed of the gas layers) 
in the 7th column 
are obtained from the best least squares fit of the
two layer radiative transfer model to the observed spectrum.
We give two infall radii in the 8th and 9th columns 
of which $\rm R_{in}(\delta V_{CS})$ is determined 
from $\delta V_{CS}$ distribution, and  $\rm R_{in}(CS)$ is determined from 
the spatial extent of the blue-skewed CS profiles
The gas ($\rm H_2$) density in 
the 10th column is the density
of a core in virial equilibrium with radius ($\rm R_{in}$) and velocity 
dispersion ($\sigma$) which approximates  the density 
associated with the gas causing the absorption dip or shoulder. 
The last column lists the mass infall rates of the cores.
For L183, all parameters are from the fit of the average spectrum over the
$\rm N_2H^+$ HM contour. As most spectra within the $\rm N_2H^+$ HM 
contour satisfy the `strong' case condition, 
we regard the L183 fit as representing the `strong' case.}  
 					       
\end{deluxetable}